\documentclass{WileyMSP-template}

\usepackage{subcaption}

\usepackage[T1]{fontenc}
\usepackage[utf8]{inputenc}
\usepackage{lmodern}
\usepackage{graphicx}

\usepackage{cite} 
\usepackage{xcolor}
\usepackage{float}
\usepackage{hyperref}
\usepackage{braket}
\usepackage{amssymb,amsmath,amsthm, array}
\usepackage{mdframed}
\usepackage{systeme,mathtools}
\usepackage{yfonts}
\usepackage{comment}

\renewcommand{\vec}[1]{\mbox{\boldmath$\mathrm{#1}$}}

\DeclareUnicodeCharacter{0301}{-}

\newcommand{\ZZ}{{\hspace{-0.75em}\phantom{\big)}}_z}

\renewcommand{\vec}[1]{\mbox{\boldmath$\mathrm{#1}$}}
\newcommand{\im}{\mathrm{i}}
\newcommand{\e}{\textrm{e}}

\DeclareMathOperator{\Tr}{Tr}

\begin{document}

\pagestyle{fancy}
%\rhead{\includegraphics[width=2.5cm]{vch-logo.png}}

\title{Quantum memory and scrambling from the perspective of a classical neural network}

\maketitle

% Author: Please give full first and last names for authors and include * after the name of all corresponding authors

\author{Dimitrios Maroulakos*}{},
\author{Andrzej Wal}{}, 
\author{Marcin Kowalik}{}, 
\author{Czes\l{}aw Jasiukiewicz}{}, 
\author{Rohit Kumar Shukla}{}, 
\author{Sunil K. Mishra},
\author{Levan Chotorlishvili}{}.

% Dedication

%\dedication{Optional dedication here. If no dedication is required, please leave blank}

% Affiliations: Please provide adacemic titles (Prof. or Dr.) for all authors where applicable, and include an institutional email address for all corresponding authors
\begin{affiliations}
Dimitrios Maroulakos\\
Doctoral School, University of Rzeszów, Rzeszów 35-310, Poland\\
Email Address: dimitriosm@dokt.ur.edu.pl

Prof. Andrzej Wal\\
Institute of Physics, Faculty of Exact and Technical Sciences, University of Rzeszów, Pigonia 1, Rzeszów 35-310, Poland

Dr. Marcin Kowalik, Prof. Czes\l{}aw Jasiukiewicz, Prof. Levan Chotorlishvili\\
Department of Physics and Medical Engineering, Rzesz\'ow University of Technology, 35-959 Rzesz\'ow, Poland

Dr. Rohit Kumar Shukla\\
Department of Chemistry, Institute of Nanotechnology and Advanced Materials; Center for Quantum Entanglement Science and Technology, Bar-Ilan University, Ramat-Gan, 52900 Israel

Prof. Sunil K. Mishra\\
Department of Physics, Indian Institute of Technology (Banaras Hindu University) Varanasi - 221005, India

\end{affiliations}

% Keywords: Please provide a minimum of three and a maximum of seven keywords, separated by commas


% Abstract should be written in the present tense and impersonal style (i.e., avoid we), and be at most 200 words long
\begin{abstract}

Entropic uncertainty relations are universal quantifiers of fundamental uncertainties of quantum measurements and are widely discussed in the quantum metrology literature. Quantum memory is a phenomenon related to the specific type of quantum correlations that allows for reducing fundamental uncertainties of quantum measurements. In the present work, the modified concept of quantum memory for time-dependent problems is proposed. We compare the time-dependent formulation of quantum memory with the out-of-time-ordered correlator (OTOC). Quantum memory is a rigorous mathematical concept that requires demanding calculations. Thus, until now, quantum memory has been discussed mainly for simple model systems and stationary problems. In the present work, we demonstrate that quantum memory can also be studied for realistic and physically relevant systems, e.g., the atomic helical spin chain, as well as the emergence and propagation of quantum correlations in time. We found that quantum memory manifests faster oscillations in time than OTOC and does not equilibrate. Furthermore, an artificial neural network is trained and asked to predict results for OTOC and quantum memory. These results show that quantum memory is more sensitive than OTOC in terms of broken inversion symmetry and the nonreciprocal effect.

\end{abstract}

\section{Introduction}\label{sec_int}

The term quantum memory has recently gained prominence, and it encompasses two significant aspects. The first aspect pertains to long-lived memory qubits, an important component of quantum networks. The critical issue of quantum technologies is the transfer of quantum states over long distances. Therefore, long-lived memory qubits and high-fidelity photon interfaces are thought to facilitate this process \cite{PRXQuantum.5.010303, PRXQuantum.5.020343, PhysRevLett.131.170802, PhysRevLett.131.033601, PhysRevLett.130.090803}. The second aspect of quantum memory is related to quantum metrology, namely advanced and improved uncertainty relations of outcomes of quantum measurements done on non-commuting operators    \cite{wang2019quantum,ming2020improved,dolatkhah2020tightening,bergh2021entanglement, PhysRevA.104.062204,chotorlishvili2019spin,song2022environment,zhu2021zero,kurashvili2022quantum, PhysRevD.103.036011}. In the present work, we are interested in the second aspect of quantum memory. Interest in the uncertainties of quantum measurements stems from Heisenberg's uncertainty principle. This principle imposes a specific, well-known limit on the precision of measurements done on a momentum $\hat p$ and a coordinate $\hat x$ of a quantum particle. In other words, Heisenberg's uncertainty principle argues that pin-point measurement of one of the variables enhances uncertainty of measurements done on the second variable and vice versa $\Delta x\Delta p\geqslant \hbar$, where $\hbar$ is  Planck's constant and violation of commutativity $\left[\hat p,\hat x\right]\neq 0 $, is essential in this context.
In the seminal work \cite{berta2010uncertainty}, the problem of quantum uncertainty was resolved through the entropic measures on a universal basis, free from the artifacts of the choice of the system's initial state. Authors of the work \cite{berta2010uncertainty} proposed the quantum memory-assisted \textit{Entropic Uncertainty Relation} (EUR) as a universal measure of uncertainties of quantum measurements. The formalism developed by the authors is relevant only to time-independent problems. In the present work, we modify this formalism for time-dependent problems relevant to quantum metrology. We propagate in time quantum memory and compare its swiftness with an out-of-time-ordered correlator (OTOC).
\begin{figure}
  \includegraphics[width=0.58\linewidth]{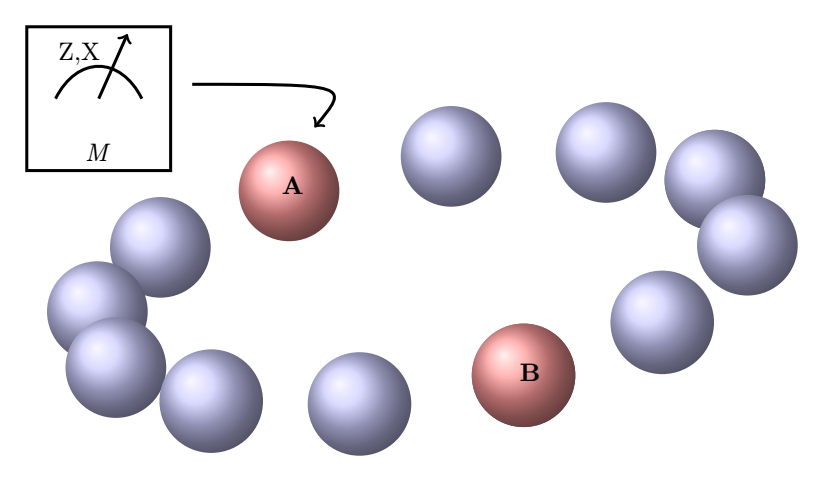} %\includegraphics[width=0.48\linewidth]{Fig1.png}
  \centering
  \caption{The proposed model mimics the experimental setup of the work \cite{Menzel2012}. The atomic chain with a length of up to 30 nm, formed on the Ir(001) surface \cite{PhysRevB.67.125422}. The spins are coupled by the nearest-neighbor ferromagnetic and the next-nearest-neighbor antiferromagnetic interactions. Competing exchange interaction terms lead to spin frustration. The Dzyaloshinskii-Moriya (DM) interaction between nearest spins stabilizes the chiral spin order. When studying the effect of quantum memory, we perform two measurements of the qubit A and measure the $ Z$ and $ X$ components of the spin. When studying OTOC, we consider two local perturbations $\hat W=\hat\sigma_A^z=\hat\sigma_1^z$ and $\hat V=\hat\sigma_B^z=\hat\sigma_{L/2}^z$, where $L$ is the length of the chain.}
  \label{schematics}
\end{figure} 
%\begin{figure}[b]
%\centering
%\includegraphics[width=0.48\linewidth]{Fig1.png}
%\caption{The proposed model mimics the experimental setup of the work \cite{Menzel2012}. The atomic chain with a length of up to 30 nm, formed on the Ir(001) surface \cite{PhysRevB.67.125422}. The spins are coupled by the nearest-neighbor ferromagnetic and the next-nearest-neighbor antiferromagnetic interactions. Competing exchange interaction terms lead to spin frustration. The Dzyaloshinskii-Moriya (DM) interaction between nearest spins stabilizes the chiral spin order. When studying the effect of quantum memory, we perform two measurements of the qubit A and measure the $ Z$ and $ X$ components of the spin. When studying OTOC, we consider two local perturbations $\hat W=\hat\sigma_A^z=\hat\sigma_1^z$ and $\hat V=\hat\sigma_B^z=\hat\sigma_{L/2}^z$, where $L$ is the length of the chain.}
%\label{schematics}
%\end{figure}
For this purpose, we study the dynamics of quantum memory in the chiral multiferroic chain, Figure \ref{schematics}. Our protocol is as follows: Suppose Alice and Bob share a tripartite system $\hat\rho_{ACB}$ where $A$ and $B$ are individual qubits at their hand, and $C$ is the chiral channel connecting them. In what follows, channel $C$ is the essence of the chiral multiferroic spin chain connecting selected two $A$ and $B$ spins in the chain. The competing nearest-neighbor ferromagnetic and next-neighbor antiferromagnetic interactions in channel $C$ lead to spin frustration and affect the transfer of quantum information from $A$ to $B$. Besides, chiral multiferroics exhibit the magnetoelectric effect, giving rise to a Dzyaloshinskii–Moriya interaction term that can be controlled by an external electric field. Dzyaloshinskii–Moriya interaction influences information transfer and formation of quantum correlations in the system.
Besides, channel $C$ is characterized by broken time and spatial inversion symmetry due to the magnetoelectric effect $\mathcal{\hat{P}}\mathcal{\hat{T}}(\hat\rho_{ACB})\neq\hat\rho_{ACB}$. Alice performs two measurements on the spin $\textbf{A}$. In the first measurement, Alice measures the $Z$ component of the spin, and in the second measurement, the $X$ component. Measurements done on $\textbf{A}$ Alice denotes by $R\equiv \lbrace Z, X\rbrace$ and stores measurements results in $\textbf{L}$. The aim of Bob is to guess the results of the measurements $R\equiv \lbrace Z, X\rbrace$ until the local perturbations due to the measurements done by Alice reach Bob through the propagation over the channel $C$. 
Bob's uncertainty about the measurement results depends on quantum memory. Measurements done by Alice are the essence of local perturbations acting on the quantum system and leading to a reshuffling of quantum correlations stored in the many-body correlated initial state \cite{Heyl2018, Heyl2018a, Heyl2013, Vosk2014}. Consequently, local perturbations influence the system's entanglement \cite{Eisert2015, Ponte2015, Azimi2014, Azimi2016}. Being the universal measure of quantum correlations, quantum memory-assisted improved EUR was never applied to time-dependent problems. We are interested in the subsequent evolution of entanglement over time and the time-dependent uncertainties of quantum measurements. For that purpose, we propose a modified version of quantum memory-assisted improved EUR and apply it to the realistic physical system, i.e., a multiferroic helical spin chain with broken inversion symmetry. A vital question is the swiftness of the spreading of quantum entanglement through the helical spin chain. The maximum rate at which correlations build in the quantum system is limited by the Lieb-Robinson bound \cite{liebrobinson}, while a quantitative criterion is provided by the OTOC of local perturbations in question. The concept of the OTOC was introduced by Larkin and Ovchinnikov\cite{larkin}, and since then, OTOC has been seen as a diagnostic tool of quantum chaos. Interest in the delocalization of quantum information (i.e., the scrambling of quantum entanglement) was renewed only recently, see~\cite{Maldacena2016, Roberts2015,Iyoda2018, Chapman2018,Swingle2017, Klug2018,Campo2017, Campisi2017,Grozdanov2018, Patel2017,Khemani2018,PhysRevB.105.224307,PhysRevA.106.022403, Rakovszky2018,Syzranov2018, Hosur2016,Halpern2017,Roberts2017,LoMonaco2023,LoMonaco2024,Touil2020,Zhuang2022} and references therein. In what follows, to explore quantum memory in the time domain, we analyze the time behavior of OTOC and quantum uncertainties. In the present work, we demonstrate that quantum memory can also be studied for realistic and physically relevant systems, e.g., the atomic helical spin chain formed on Ir(001) surface. To analyze the predictability of quantum memory and OTOC, we exploit the artificial neural network method and explore the time series of the data obtained for particular sets of parameters. We note that our approach is based on the deep neural network backpropagation algorithm, and the architecture of our network includes several hidden layers of classical neurons \cite{nielsen2015} and not quantum machine learning and quantum perceptrons \cite{beer2020training,cerezo2021cost, PRXQuantum.3.030101, PhysRevX.11.041011}. We will show that predictions of the neural network for quantum memory drastically depend on the Dzyaloshinskii-Moriya interaction and broken inversion symmetry in the system. Namely, the difference between the exact and predicted values of improved EUR depends on the strength of the Dzyaloshinskii-Moriya interaction. 
The work is organized as follows: In section \textbf{2}, we describe the model in question, a helical spin chain with broken inversion symmetry, and the methods used in our study to quantify the time-dependent quantum correlations in the system (quantum memory and OTOC). We also briefly describe the neural network architecture used in the learning procedure. In section \textbf{3} we present the results of the time propagation of quantum correlations in the system quantified in terms of OTOC and quantum memory. We also present OTOC and quantum memory results predicted by the neural network and compare predicted results with exact results. We conclude the work in section \textbf{4}.

\section{Methods}
\subsection{Model}
\label{sec_model}

We aim to explore the time-dependent quantum memory and scrambling process in the oxide-based, spin-driven chiral multiferroic (MF) systems,  LiCu$_2$O$_2$ could serve as a prototype material. The remarkable experimental properties of MF systems were uncovered in works \cite{Wang2003, Ramesh2007, Bibes2008, Fiebig2005, Hemberger2007, Menzel2012, Stagraczynski2017,Khomeriki2016, PhysRevB.91.041408}. The dual ferroelectric - ferromagnetic order parameters make MFs especially attractive for spintronics. The strong magnetoelectric (ME) coupling allows one to control the magnetization dynamics and ground state magnetic order using an external electric field. This option is a positive prospect for developing environmentally friendly nano-devices and for technological applications. The Hamiltonian of the chiral MF system reads:  
\begin{eqnarray}\label{Hamiltonian}
&&\hat{H} = J_1 \sum^{L}_{i=1}\hat{\vec{S}}_i \cdot \hat{\vec{S}}_{i+1} + J_2 \sum^L_{i=1}\hat{\vec{S}}_i\cdot \hat{\vec{S}}_{i+2} + D \sum^{L}_{i=1} \left(\hat{\vec{S}}_{i} \times \hat{\vec{S}}_{i+1} \right)\ZZ\,.
\end{eqnarray}
The $L$ effective {spin}-$1/2$ moments localized at sites $i$ are described by the operators $\mathbf{\hat S}_i$. The nearest neighbor exchange interaction is ferromagnetic ${J_1<0}$, while the next nearest neighbor term is antiferromagnetic ${J_2>0}$ and leads to a frustrated spin order. The last term in Eq.\eqref{Hamiltonian} describes coupling of the ferroelectric polarization with the applied external electric field ${-{\bf P}\cdot{\bf E}}=D \sum^{L}_{i=1} (\hat{\vec{S}}_{i} \times \hat{\vec{S}}_{i+1} )_z$ and $D=E_y g^{\phantom{\dagger}}_{\mathrm{ME}}$. The ferroelectric polarization ${{\bf P}=g^{\phantom{\dagger}}_{\mathrm{ME}}(\mathbf{\hat{e}_{ij}} \times \mathbf{\hat{\varkappa}})}$ can be expressed in terms of the $z$ component of the vector chirality $\hat{\varkappa}^{z} = \sum^L_{i=1}\hat{\varkappa}_{i}^{z} = \sum^L_{i=1} (\mathbf{\hat S}_i \times \mathbf{\hat S}_{i+1})_z$, where
$\mathbf{\hat{e}_{ij}}$ is the unit vector connecting spins, and the ME coupling constant $g^{\phantom{\dagger}}_{\mathrm{ME}}$. The chirality ${\varkappa^{z}=\langle \hat{\varkappa}^{z} \rangle}$ is the order parameter that is non-zero in the chiral phase and disappears in the case of collinear magnetic order \cite{Cheong2007,Katsura2005,Mostovoy2006,PhysRevB.111.024315,  Sergienko2006,Tokura2014,kolesnikov2025magnetic,kolesnikov2025magnetization}. The constant $D$ combines the effect of the electric field with the ME coupling and mimics the Dzyaloshinskii-Moriya (DM) interaction term \cite{Moon2013}. In the present work, we explore the impact of chirality on the delocalization process of quantum information. A unitary local rotation of spins $\hat{S}_{j}^{+} \rightarrow \hat{S}_{j}^{+}e^{-\im \Theta j}$, $\hat{S}_{j}^{-}\rightarrow \hat{S}_{j}^{-}e^{\im \Theta j}$ around the $z$ axis by the angle $\Theta =-\arctan\left(D/J_{1}\right)$ converts the Hamiltonian to
 \begin{eqnarray}\label{Hamiltonian2}
  \hat{H_T} &=& J_1 \sum^L_{i=1} \hat{S}_i^{z} \hat{S}_{i+1}^{z} + \frac{J^\prime_1}{2} \sum^L_{i=1}\left(\hat{S}_i^{+} \hat{S}_{i+1}^{-}+\hat{S}_i^{-} \hat{S}_{i+1}^{+}\right)\nonumber\\
  &+&J_2 \sum^L_{i=1}\hat{S}_i^{z} \hat{S}_{i+2}^{z} + \frac{J^\prime_2}{2} \sum^L_{i=1} \left(\hat{S}_i^{+} \hat{S}_{i+2}^{-}+\hat{S}_i^{-} \hat{S}_{i+2}^{+}\right)\nonumber\\
   &-& B^z\sum^L_{i=1}  \hat{S}_i^z  - D^\prime \sum^L_{i=1}\left(\hat{\vec{S}}_{i} \times \hat{\vec{S}}_{i+1} \right)\ZZ.
\end{eqnarray}
Here  $J^\prime_1 = \sqrt{J_1^{2}+D^{2}}$, $J^\prime_{2}=J_{2}\left(J_{1}^{2}-D^{2}\right)/\left(J_{1}^{2}+D^{2}\right)$, and $D^\prime=DJ_{1}J_{2}/\left(J_{1}^{2}+D^{2}\right)$. Depending on the values of the parameters, the ground state of \eqref{Hamiltonian2} can be either ferromagnetic, chiral, or nematic \cite{Azimi2014,Azimi2016}.

\noindent It is easy to see that the total $\hat{S}^{z}=\sum_{n}{S}^{z}_{n}$ is conserved $[\hat{S}^{z},\hat{H}]=0$ and therefore the Hamiltonian can be put into a block-diagonal structure. This allows one to solve the single magnon sector exactly. Taking the eigenstates of the total momentum as an ansatz for the solution $|\Psi_k\rangle=\frac{1}{\sqrt{L}}\sum_j \mathrm{e}^{-\im k j}|j\rangle$ we diagonalize the Hamiltonian \eqref{Hamiltonian} and deduce the following dispersion relations
\begin{equation}\label{dispersionrelations}
 \epsilon_{\pm}(k) = J_1 \cos k + J_2 \cos 2k \pm D \sin k .
\end{equation}
Taking into account finite length $L$ chain and quantization condition $k=\frac{2\pi}{L}n$ we specify the eigenstates and spectrum with $\hbar=1$: 
\begin{eqnarray}\label{eigenstates}
&&\omega_n^{\pm}= J_1 \cos(2\pi n/L) + J_2 \cos (4\pi n/L) \pm D \sin (2\pi n/L), 
\end{eqnarray}
\begin{eqnarray}
&&\ket{\varphi_n}=\frac{1}{\sqrt{L}}\sum\limits_{j=1}^L\, \mathrm{e}^{-\im 2\pi nj/L}\ket{j}.
\end{eqnarray}
For the mismatch between the left/right group velocities in this single magnon sector, we deduce that $\Delta v_{g}(k)=v_{g}^{+}(k)-v_{g}^{-}(k)=2D\cos(k)$. The next to nearest neighbor term (${J_{2}\neq0}$) leads to an extra peak in the dispersion relation and two pairs of non-equal peaks in the left and right branches of the group velocity. The later directly influences the asymmetry of the scrambling and time-dependent quantum memory.
Apparently, the group velocity shows a certain asymmetry for left/right propagating spin waves $v_{g}^{\pm}(k)={\partial\omega_{\pm}(k)}/{\partial k}$. Therefore, we expect to see an impact of the quantum Doppler effect on quantum memory and scrambling.

\subsection{Quantum memory, definition and formalism}\label{sec_memory}

We adopt the periodic boundary conditions and assume that the qubits of Alice  $\textbf{A}$ and Bob $\textbf{B}$ are located at the diameter of the chiral spin ring. We set the length of the chain equal to hundred spins $L=100$. Thus the distance between selected qubits $\textbf{A}$ and $\textbf{B}$ is equal to $L/2$. The total density matrix of the system is given by $\hat\rho_{ABC}(t)=\hat U^{-1}(t)\hat\rho_{ABC}(0)\hat U(t)$, where $\hat U(t)$ is the evolution operator and $\hat\rho_{ABC}(0)=\ket{\psi}\bra{\psi}_{ABC}$ is the initial state. After tracing out of the chiral chain, we construct the reduced density matrix $\hat\varrho_{AB}(t)=\text{Tr}_C\left(\hat\rho_{ABC}(t)\right)$. Then, the reduced density matrices of Alice and Bob's subsystems are defined as follows: $\hat\varrho_A(t)=Tr_B\left(\hat\varrho_{AB}(t)\right)$ and $\hat\varrho_B(t)=Tr_A\left(\hat\varrho_{AB}(t)\right)$. The von Neumann Entropy of the entire system is given by  $S(AB)=-Tr\left(\hat\varrho_{AB}\log\hat\varrho_{AB}\right)$. The conditional quantum entropy has a form $S(A\vert B)=S(AB)-S(B)$. After Alice performs two measurements and measures $z$ and $x$ components of her qubit \textbf{A}, the subsequent post-measurement density matrices are given by
%\begin{widetext}
\begin{eqnarray}
\label{eq:rhoZ}
&&\hat\varrho_{Z,AB}(t)=\sum\limits_{n}\ket{\psi_n}\bra{\psi_n}_A\otimes Tr_A\left\lbrace(\ket{\psi_n}\bra{\psi_n}_A\otimes\hat I_B)\hat\varrho_{AB}(t)\right\rbrace,\\
\label{eq:rhoX}
&&\hat\varrho_{X,AB}(t)=\sum\limits_{n}\ket{\phi_n}\bra{\phi_n}_A\otimes Tr_A\left\lbrace(\ket{\phi_n}\bra{\phi_n}_A\otimes\hat I_B)\hat\varrho_{AB}(t)\right\rbrace.
\end{eqnarray}
%\end{widetext}
Here $\ket{\psi_{1,2}}\equiv \ket{0}_A,\,\ket{1}_A$ and $\ket{\phi_{1,2}}=\frac{1}{\sqrt{2}}\left(\ket{0}_A\pm\ket{1}_A\right)$ are eigen functions of the $z$, $x$ components of the qubit $\textbf{A}$.
 Alice performs two measurements on her particle   \textbf{A} and broadcasts her measurement choice to Bob. Bob wants to guess Alice's outcome precisely by measuring his particle B with the help of the received classical information (i.e., Alice's choice of measurement). Bob's ignorance about Alice's measurements is given by EUR \cite{berta2010uncertainty,wilde2013,PhysRevB.108.134411}:
\begin{eqnarray}
\label{eq:BobsIgnorance}
 S(X\vert B)+S(Z\vert B)\geqslant \log_2(1/c)+S(A\vert B),\\
\label{eq:cij}
 c=\max\left\lbrace \vert\bra{\psi_i}\ket{\phi_j}\vert^2\right\rbrace. 
\end{eqnarray}
Here $c=1/2$ is a measure of complementarity and $S(X\vert B)$, $S(Z\vert B)$ are conditional quantum entropies of the states (\ref{eq:rhoZ}), (\ref{eq:rhoX}). \vspace{0.2cm}\\
The meaning of (\ref{eq:BobsIgnorance}), (\ref{eq:cij})
is as follows: The left-hand side of (\ref{eq:BobsIgnorance}) defines the uncertainty about measurement results
(two measurements done on the $\textbf{x}$ and $\textbf{z}$ spin components of  the particle \textbf{A}). The right-hand side of (\ref{eq:BobsIgnorance}) defines the lower bound of this uncertainty. The first term on the right-hand side $\log_2(1/c)$ is positive. However, conditional quantum entropy $S(A\vert B)$ can be negative for entangled states. Dependence of the lower bound of uncertainties on the correlation between the subsystems $\textbf{A}$ and $\textbf{B}$ is related to the effect of quantum memory \cite{PhysRevA.102.012206,PhysRevA.102.052227,PhysRevA.106.062219,PhysRevA.108.L050202}. 
Negative conditional quantum entropy $S(A\vert B)$ means that the state $\hat\varrho_{AB}$ for sure is entangled, but the converse is not always true. Quantum memory can reduce the right-hand side of (\ref{eq:cij}), which permits measurement uncertainties quantified by the left-hand side to take a smaller value. 

\subsection{OTOC, definition and formalism}\label{sec_OTOC}

Let us consider two unitary operators $\hat{V}$ and $\hat{W}$ describing local perturbations to the system and their unitary time evolution under a Hamiltonian $\hat{H}$ which we will specify below, ${\hat{V}\left(t\right)=\exp (\im\hat{H}t)\,\hat{V}\exp(-\im\hat{H}t)}$, and similarly for ${\hat{W}\left(t\right)}$. Here, we measure time such that $\hbar=1$. Then, the OTOC is defined as \cite{PhysRevB.99.184202,PhysRevB.99.224305,PhysRevB.105.104202,PhysRevB.107.L220203,shukla2023quantum}:
\begin{equation}\label{OTOC1}
C\left(t\right)= \left\langle\left[\hat{W}(t),\hat{V}\right]^{\dag}\left[\hat{W}(t),\hat{V}\right]\right\rangle\,,
\end{equation}
or in an alternative but equivalent form as $C\left(t\right)=2-2\Re F\left(t\right)$ where $F\left(t\right)=\left\langle \hat{W}(t)^{\dag}\hat{V}^{\dag}\hat{W}(t)\hat{V}\right\rangle$. Here parentheses $\langle\ldots\rangle$, if not otherwise specified, denote either the quantum mechanical ground state average $\langle\ldots\rangle$=$\langle\psi|\ldots|\psi\rangle$, or the finite temperature thermal average $\langle\ldots\rangle=Z^{-1}\text{Tr}\left(\e^{-\beta \hat{H}}\ldots\right)$ with inverse temperature ${\beta=1/T}$ and the Boltzmann constant scaled to ${k_B=1}$. At the initial moment, as follows from the definition, the OTOC is zero ${C(0)=0}$, provided ${[\hat W,\hat V]=0}$.
OTOC demonstrates several interesting physical features. For example, OTOC points to dynamical phase transitions\cite{Heyl2018a,PhysRevB.107.L100419}. Disorder slows the growth of  $C\left(t\right)$ in time, and therefore scrambling can be used to identify the many-body localization phase\cite{Swingle2017}. Scrambling itself is nothing other than the zero velocity Lieb Robinson bound\cite{Hamza2012}
\begin{equation}\label{OTOC2}
C\left(t\right)=\|\left[W(t),V\right]\|=\min\left(|t|,1\right)\mathrm{e}^{-\eta d(\hat{W},\hat{V})}\,.
\end{equation}

\noindent In the sense that a zero Lieb–Robinson velocity reflects either the absence of propagation of correlations or extremely slow propagation caused by the disorder \cite{goihl2019experimentally}. In this context, ${\eta=\mathrm{const}}$ represents a constant value, $d(\hat{W}, \hat{V})$ denotes the distance between operators on the lattice (for example, spin operators $\exp (\hat{S}_{n}^{z})$ of a spin chain), and the Frobenius norm of the unitary operator $\hat{A}$ is defined as $\|\hat{A}\|=\text{Tr}(\hat{A}^{\dag}\hat{A})$. Rather interesting and enlightening is a semi-classical interpretation of scrambling. For canonical momentum and coordinate operators ${V\equiv P}$, ${W(t)\equiv q(t)}$ one deduces the equation, valid on the short time scale \cite{Maldacena2016}, $C\left(t\right)=\hbar^{2}\exp\left(2\lambda^{\pm}_{L}t\right)$.  The scrambling time is specified in terms of the classical Lyapunov exponent $\lambda_{L}$ and is equal to the Ehrenfest time $\tau\approx\frac{1}{\lambda_{L}}\ln1/\hbar$. On the other hand, a purely quantum analysis shows that the radius of the operator linearly increases in time,\cite{Roberts2015} independent of whether the quantum system is integrable or chaotic \cite{Roberts2015}. This means that in the qubit system (for example, a Heisenberg spin chain), the time required for the formation of correlations between initially commuting operators $\left[\sigma_{n}^{\alpha},\sigma_{m}^{\beta}\right]=2\im\delta_{nm}\epsilon^{\alpha\beta\gamma}\sigma^{\gamma}$ will increase linearly with the distance ${d=|n-m|}$ between them, and hence $\left[\sigma_{n}^{\alpha}\left(t\right),\sigma_{m}^{\beta}\right]\neq 0$, for ${n\neq m}$. Note that customarily scrambling is an irreversible process after entanglement is spread across the system; it cannot be unscrambled\cite{Campisi2017}. 
We use the canonical four-point OTOC widely adopted in the literature as the minimal scrambling diagnostic tool. In the recent literature, higher-point k-OTO correlators are discussed, which are beyond the scope of the present study \cite{Bagrets2017,Kawamoto2025,Anous2019,Haehl2021,Iaconis2021,Roberts2017,Haehl2019,Fujii2025}. However, we note that even higher-point k-OTO correlators still remain operator-based scrambling diagnostics, while Quantum Memory is a state-based entropic quantity. OTOC can be interpreted as two wave functions that evolved in time differently \cite{Singh2022}: 
Let $|\psi(0)\rangle$ be the initial pure state wave function, which is time evolved in the following steps: first, it is perturbed at ${t=0}$ with a local unitary operator $\hat{V}$, then evolved forward under the unitary evolution operator ${\hat{U}=\exp (-\im\hat{H}t)}$ until ${t=\tau}$, it is then perturbed at ${t=\tau}$ with a local unitary operator $\hat{W}$, and evolved backward from ${t=\tau}$ to ${t=2\tau}$ under $\hat{U^{\dagger}}$. Hence the time evolved wave function is $|\psi(2\tau)\rangle =\hat{U^{\dagger}}\hat{W}\hat{U}\hat{V}|\psi(0)\rangle=\hat{W}(t)\hat{V}|\psi(0)\rangle$. For the second wave function, the order of the applied perturbations is permuted, i.e.~first $\hat{W}$ at ${t=\tau}$ and then $\hat{V}$ at ${t=2\tau}$. Therefore the second wavefunction is $|\phi(2\tau)\rangle=\hat{V}\hat{U^{\dagger}}\hat{W}\hat{U}|\psi(0)\rangle=\hat{V}\hat{W}(t)|\psi(0)\rangle$ and their overlap is equivalent to the OTOC $F(t)=\langle\phi(t)|\psi(t)\rangle$. What breaks the time inversion symmetry for the OTOC is the permuted sequence of operators $\hat{W}$ and $\hat{V}$. However in lattice spin systems with preserved spatial inversion symmetry $\mathcal{\hat{P}}\hat{H}=\hat{H}$, the spatial inversion ${\mathcal{\hat{P}}d(\hat{W}, \hat{V})}={-d(\hat{W}, \hat{V})}={d(\hat{V}, \hat{W})}$ can restore the permuted order between $\hat{V}$, $\hat{W}$.  Permuting just a single wavefunction one finds $C(t)=2-2\Re(\langle\phi(t)|\mathcal{\hat{P}}\mathcal{\hat{T}}|\psi(t)\rangle)=C(0)$. Thus, scrambled quantum entanglement formally can be unscrambled by spatial inversion. However in the chiral systems ${\mathcal{\hat{P}}\hat{H}\neq\hat{H}}$ unscrambling procedure fails. In this case OTOC is $\mathcal{\hat{P}}\mathcal{\hat{T}}$ invariant $\mathcal{\hat{P}}\mathcal{\hat{T}}C(t)=C(t)$ only if the initial state of the system $|\psi(0)\rangle$ is $\mathcal{\hat{P}}\mathcal{\hat{T}}$ invariant $\mathcal{\hat{P}}\mathcal{\hat{T}}|\psi(0)\rangle=|\psi(0)\rangle$. Indeed after applying the spatial inversion operator we deduce: $\mathcal{\hat{P}}|\psi(2\tau)\rangle=\hat{V}(t)\hat{W}|\psi(0)\rangle$, $\mathcal{\hat{P}}|\phi(2\tau)\rangle=\hat{W}\hat{V}(t)|\psi(0)\rangle$ while after time inversion we deduce $\mathcal{\hat{T}}\mathcal{\hat{P}}|\psi(2\tau)\rangle=\hat{V}(-t)\hat{W}|\psi(0)\rangle$ and $\mathcal{\hat{T}}\mathcal{\hat{P}}|\phi(2\tau)\rangle=\hat{W}\hat{V}(-t)|\psi(0)\rangle$. We then obtain: $\mathcal{\hat{T}}\mathcal{\hat{P}}F(t)=\langle\psi|\phi\rangle=F^{\ast}(t)$ meaning that $\Re \mathcal{\hat{T}}\mathcal{\hat{P}}F\left(t\right)=\Re F\left(t\right)$.
For further convenience, in the four-operator product forms $\hat{U^{\dagger}}\hat{W}\hat{U}\hat{V}$, between the operators we insert identity operators using the completeness of the eigen-states $\mathcal{I}=\sum\limits_n\ket{n}\bra{n}$. Then, considering local perturbations $\hat W=\hat\sigma_1^z$, $\hat V=\hat\sigma_{L/2}^z$ and the spectral decomposition of the unitary evolution operator, we deduce:
\begin{eqnarray}\label{wavefunctionone}
&&C^\pm(t)=1-\text{Re}\left[\sum\limits_{n_2,n_3,n_4} \mathrm{e}^{\im  t(E^{\pm}_{n_1}-E^{\pm}_{n_2}+E^{\pm}_{n_3}-E^{\pm}_{n_4})}\mathcal{A}_{n_1n_2}^{n_3n_4}\right],\nonumber\\
 &&\mathcal{A}_{n_1n_2}^{n_3n_4}=\bra{n_1}\hat\sigma_{L/2}^z\ket{n_2}\cdot
\bra{n_2}\hat\sigma_{1}^z\ket{n_3}\cdot\bra{n_3}\hat\sigma_{L/2}^z\ket{n_4}\cdot\bra{n_4}\hat\sigma_{1}^z\ket{n_1}.
\end{eqnarray}
In the infinite time limit, OTOC relaxes to the time-independent value \cite{PhysRevB.104.104306,PhysRevLett.123.010601}:
\begin{eqnarray}\label{OTOC relaxes}
&&C^\pm(\infty)=1-\mathcal{F},\nonumber\\
&&\mathcal{F}=\sum\limits_p(\hat\sigma_1)^2_{pp}(\hat\sigma_{L/2})_{pp}^2\nonumber\\
&& + \sum\limits_{p,q\neq p}(\hat\sigma_1)_{pp}(\hat\sigma_{L/2})_{pq}(\hat\sigma_1)_{qq}(\hat\sigma_{L/2})_{qp}\nonumber\\
&& + \sum\limits_{p,q\neq p}(\hat\sigma_1)_{pq}(\hat\sigma_{L/2})_{qq}(\hat\sigma_1)_{qp}(\hat\sigma_{L/2})_{pp}.
\end{eqnarray}
However, here we are interested in short-period dynamics. 

\subsection{Artificial Neuron}
\begin{figure}[ht]
\centering
\includegraphics[width=0.3\linewidth]{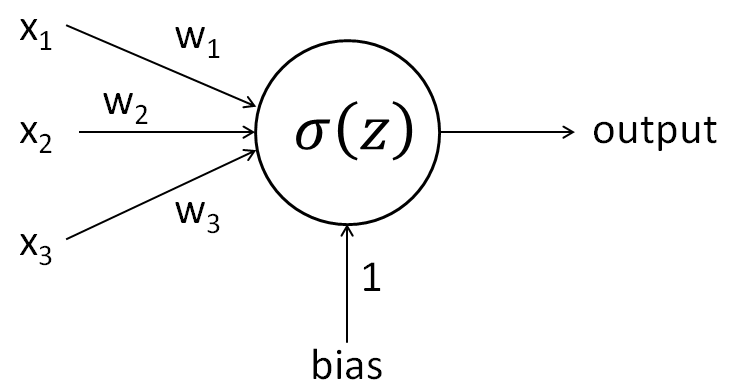}
\caption{The pictorial plot of the artificial neurons which are used to predict the time-dependence of OTOC and quantum memory.}
\label{fig:artificial_neuron}
\end{figure}

We exploit neural networks to compare exact and predicted values of OTOC and quantum memory via the analysis of the time series of quantities in question obtained for different sets of parameters. The main interest concerns DM interaction and broken inversion symmetry. We will study whether DM interaction can influence the accuracy of network predictions for quantum memory. 
An \textit{artificial neuron} (Figure \ref{fig:artificial_neuron}), also known as a perceptron \cite{rosenblatt1958perceptron}, is a fundamental building block of artificial neural networks (ANNs). Each neuron receives a set of input signals \( x_1, x_2, \ldots, x_n \), multiplied by an associated weight \( w_1, w_2, \ldots, w_n \). The weighted inputs are summed and added to a bias term \( b \), resulting in the expression:

\begin{equation}
z = \sum_{i=1}^{n} w_i x_i + b \, .
\end{equation}

\noindent This value \( z \) is then passed through a \textit{nonlinear activation function} \( \sigma \), and produces the neuron's output:

\begin{equation}
o = \sigma(z) = \sigma\left(\sum_{i=1}^{n} w_i x_i + b\right) .
\end{equation}

\noindent The activation function brings nonlinearity into the model and allows the network to learn complex patterns. The simplest activation functions are the \textit{binary step function} and \textit{sigmoid function}. Both functions map the input to a range between $0$ and $1$. In this range, output values near $0$ can be interpreted as the neuron being "inactive," while values near $1$ indicate "activation". The most commonly used activation function is ReLU \cite{Hinton_ReLU}, which has the form of

\begin{equation}
\text{ReLU}(x) = \max(0, x) .
\end{equation}

\noindent This function outputs the input directly if it is positive; otherwise, it outputs zero. ReLU introduces nonlinearity while being computationally efficient and helps mitigate the vanishing gradient problem often encountered with sigmoid or tanh functions.

\subsection{Artificial Neural Networks (ANNs)}
\begin{figure}[ht]
\centering
\includegraphics[width=0.48\linewidth]{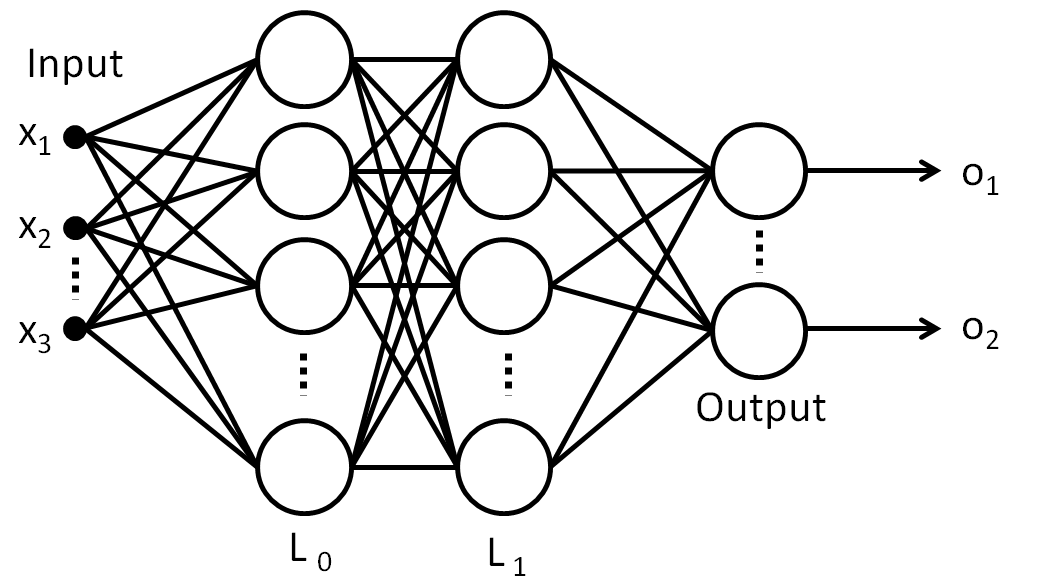}
\caption{Schematic of an ANN used to predict OTOC  $C(t)$ and quantum memory, for given values of exchange constants $J_1$, $J_2$, and DM constant $D$. The ANN consists of two hidden dense layers $L_0$ and $L_1$ with 64 and 32 neurons, respectively. The output layer contains 25 neurons.}
\label{ANN_architecture}
\end{figure}
Artificial neurons can be connected to each other to form a \textit{neural network}, either in a structured or unstructured configuration\cite{nielsen2015}. A typical architecture of an artificial neural network comprises several \textit{layers}. The neurons in a given layer are not connected within the layer, but are connected to neurons from the adjacent layer. Each layer receives inputs from the previous layer and transmits outputs to the next.
In a \textit{feedforward neural network}, information flows in one direction from the input layer to the output layer. The operation of each layer \( L \) can be expressed as:
\begin{equation}
\text{Out}_L = \sigma(\text{Out}_{L-1} \cdot W_L + B_L),
\end{equation}
where $\text{Out}_{L-1}$ is the output vector or matrix in case of using batches of data of layer $L-1$, $W_L$ is the weight matrix, $B_L$ is the bias vector associated with layer $L$, and $\sigma$ is the activation function applied element-wise.

\subsection{Learning in Neural Networks}

Artificial neural networks form a class of machine learning algorithms that aim to approximate a function $y(x)$ based on the data $x$. The \textit{learning process} involves adjusting the weights and biases so that the network's output closely matches the expected output \( y \). A \textit{cost function} measures the network's performance. Widly used cost function is the \textit{Mean Squared Error (MSE)}:
\begin{equation}
\text{MSE}(w, b) = \frac{1}{n} \sum_{i=1}^{n} \left\|Out(x_i; w, b) - y_i\right\|^2.
\end{equation}
Here, $w$ and $b$ denote all weights and biases in the network, $n$ is the number of training samples, $x_i$ is the input, and $y_i$  is the true output. The output $\text{Out}$ depends on $x$, $w$ and $b$.

\noindent To minimize the cost function, the \textit{gradient} of the cost function   
concerning each weight and bias is computed using \textit{backpropagation} algorithm \cite{Rumelhart, Rojas1996}. 
Backpropagation allows the network to iteratively adjust these parameters via \textit{gradient descent} or related optimization algorithms such as ADAM \cite{AdamOptimizator}. A critical requirement for this process is that the cost function must be differentiable with respect to the parameters. The update rules for the weights $w$ and biases $b$ using standard gradient descent are given by:
\begin{align}
w_k &\rightarrow w_k' = w_k - \eta \frac{\partial C}{\partial w_k} \label{eq:weight_update} ,\\
b_l &\rightarrow b_l' = b_l - \eta \frac{\partial C}{\partial b_l} \label{eq:bias_update},
\end{align}
\noindent where $\eta$ denotes the learning rate and $C$ is the cost function.(\ref{eq:weight_update}), (\ref{eq:bias_update}) illustrate how each parameter is updated in the direction that minimizes the cost.
Neural networks can be trained for \textit{regression tasks}  (predicting continuous values) or \textit{classification tasks}  (predicting discrete labels). The choice of cost function and network architecture depends on the specific problem being addressed.

\section{Results and Discussion}

\subsection{Quantum memory in single excitation basis}

At first, we consider the single excitation problem that allows us to obtain analytic results for the spin chain of arbitrary length. Taking into account (\ref{eigenstates}) we derive the analytic formula for the evolved in time density matrix $\hat\rho_{ABC}(t)=e^{-i\hat H t}\hat\rho_{ABC}(0)e^{i\hat H t}$, $\hat\rho_{ABC}(0)=\ket{\psi_{ABC}}\bra{\psi_{ABC}}$ as follows:
%\begin{widetext}
\begin{eqnarray}\label{the analytic formula}
\hat\rho_{ABC}(t)&=&\frac{1}{L^2}\sum\limits_{n,m=1}^L\sum\limits_{j,j'=1}^L\sum\limits_{q,q'=1}^L \mathrm{e}^{\im (\omega_n^{\pm}-\omega_m^{\pm})t} \, \mathrm{e}^{\frac{2\im \pi}{L}(n(q-q')+m(j'-j))} \times \\ \nonumber
&&\times  \langle\psi_{ABC}\vert q'\rangle\langle j'\vert\psi_{ABC}\rangle\ket{j}\bra{q}.
\end{eqnarray}
%\end{widetext}
Taking into account (\ref{the analytic formula}) and (\ref{eq:rhoZ})-(\ref{eq:cij}), we calculate quantum memory propagating clockwise and anti-clockwise directions. 
For further convenience, we rewrite (\ref{the analytic formula}) in the following compact form $\hat\rho_{ABC}^{\pm}(t)=\sum\limits_{j,q=1}^L\rho_{jq}^{\pm}\ket{j}\bra{q}$, where we introduced the notation of matrix elements 
\begin{eqnarray}
\rho_{jq}^{\pm}&=&\frac{1}{L^2}\sum\limits_{n,m=1}^L\sum\limits_{j'=1}^L\sum\limits_{q'=1}^L \mathrm{e}^{\im (\omega_n^{\pm}-\omega_m^{\pm})t} \mathrm{e}^{\frac{2\im \pi}{L}(n(q-q')+m(j'-j))}\times \\  \nonumber
&&\times  \langle\psi_{ABC}\vert q'\rangle\langle j'\vert\psi_{ABC}\rangle. \nonumber
\end{eqnarray}
%$$\rho_{jq}^{\pm}=\frac{1}{L^2}\sum\limits_{n,m=1}^L\sum\limits_{j'=1}^L\sum\limits_{q'=1}^L \mathrm{e}^{\im (\omega_n^{\pm}-\omega_m^{\pm})t} \mathrm{e}^{\frac{2\im \pi}{L}(n(q-q')+m(j'-j))}\times\\\langle\psi_{ABC}\vert q'\rangle\langle j'\vert\psi_{ABC}\rangle.$$ 
When the system initially is prepared in the ground state $\ket{\psi_{ABC}}=\ket{1_1\ldots 0_L}$, coefficients take a simpler form $\rho_{jq}^{\pm}=\frac{1}{L^2}\sum\limits_{n,m=1}^L \, \mathrm{e}^{\im (\omega_n^{\pm}-\omega_m^{\pm})t} \mathrm{e}^{\frac{2\im \pi}{L}(n(q-1)+m(1-j))}$. 
We present the density matrix in the form
\begin{eqnarray}\label{we present the density matrix in the form}
\hat\rho_{ABC}^{\pm}(t)&=&\rho^\pm_{11}\ket{1}\bra{1}_A\otimes\left(\ket{0}\bra{0}\right)_C^{\otimes (L-2)}\otimes\ket{0}\bra{0}_B\nonumber\\
&&+\rho_{LL}^\pm\ket{0}\bra{0}_A\otimes\left(\ket{0}\bra{0}\right)_C^{\otimes (L-2)}\ket{1}\bra{1}_B\nonumber\\
&&+\ket{0}\bra{0}_A\otimes\ket{0}\bra{0}_B\otimes\sum\limits_{j,q=2}^{L-1}\rho^\pm_{jq}\ket{j}\bra{q}.
\end{eqnarray}
Then the reduced density matrix, after tracing the channel part, takes the form:
\begin{eqnarray}\label{reduced density matrix takes the form}
\hat\rho_{AB}^\pm &=&\ket{1}\bra{1}_A\otimes\ket{0}\bra{0}_B\rho_{11}^\pm+\ket{0}\bra{0}_A\otimes\ket{1}\bra{1}_B\rho_{LL}^\pm\nonumber\\
&&+\ket{0}\bra{0}_A\otimes\ket{0}\bra{0}_B\sum\limits_{j=2}^{L-1}\rho_{jj}^\pm.
\end{eqnarray}
The explicit expressions of the post-measurement density matrices read:
\begin{eqnarray}\label{reducedB}
&&\hat\rho_B=\text{Tr}_A\left(\hat\rho_{AB}^\pm\right)=\ket{0}\bra{0}_B\left(\rho_{11}^\pm+\sum\limits_{j=2}^{L-1}\rho_{jj}^\pm\right)+\ket{1}\bra{1}_B\rho_{LL}^\pm  \, ,
\end{eqnarray}
\begin{eqnarray}
\label{eq:ExplicitrhoZ}
\hat\varrho^\pm_{Z,AB}(t)&=&\ket{0}\bra{0}_A\otimes \left(\ket{1}\bra{1}_B\rho_{LL}^\pm+\ket{0}\bra{0}_B\sum\limits_{j=2}^{L-1}\rho_{jj}^\pm\right)\nonumber\\
&& +\ket{1}\bra{1}_A\otimes\ket{0}\bra{0}_B\rho_{11}^\pm,
\end{eqnarray}
\begin{eqnarray}\label{eq:ExplicitrhoX}
\hat\varrho^\pm_{X,AB}(t)&=&\frac{1}{2}\big(\ket{0}\bra{0}_A\otimes\ket{0}\bra{0}_B+\ket{1}\bra{1}_A\otimes\ket{0}\bra{0}_B\big)\rho_{11}^\pm\nonumber\\
&&+\frac{1}{2}\big(\ket{0}\bra{0}_A\otimes\ket{1}\bra{1}_B+\ket{1}\bra{1}_A\otimes\ket{1}\bra{1}_B\big)\rho_{LL}^\pm\nonumber\\
&&+\frac{1}{2}\big(\ket{0}\bra{0}_A\otimes\ket{0}\bra{0}_B+\ket{1}\bra{1}_A\otimes\ket{0}\bra{0}_B\big)\sum\limits_{j=2}^{L-1}\rho_{jj}^\pm.
\end{eqnarray}
Taking into account (\ref{reduced density matrix takes the form})-(\ref{eq:ExplicitrhoX}) we calculate analytic expressions of entropies:
\begin{eqnarray}\label{analytic expressions of entropies}
&& S^\pm(X\vert B)=S^{\pm}(\hat\rho_{X,AB}^\pm)-S^{\pm}(\hat\rho^\pm_{X,B}),\\
&& S^\pm(Z\vert B)=S^{\pm}(\hat\rho_{Z,AB}^\pm)-S^{\pm}(\hat\rho_{Z,B}^\pm),\\
&& S^\pm(A\vert B)=S^{\pm}(\hat\rho_{AB}^\pm)-S^{\pm}(\hat\rho_B^\pm).
\end{eqnarray}
Taking into account spectral decomposition of the operator over the eigenstate basis $\log\hat A=\sum\limits_n \ket{n}\bra{n}\log(A_n)$, the explicit expressions of the entropies read:
\begin{eqnarray}\label{XBexplicit}
S^\pm(X\vert B)=1,  
\end{eqnarray}
\begin{eqnarray}\label{ZBexplicit}
S^\pm(Z\vert B)&=&-\rho^\pm_{11}\log\rho^\pm_{11}-\sum\limits_{j=2}^{L-1}\rho_{jj}^\pm\log\left(\sum\limits_{i=2}^{L-1}\rho_{ii}^\pm\right)\nonumber\\
&&+\sum\limits_{j=1}^{L-1}\rho_{jj}^\pm\log\left(\sum\limits_{i=1}^{L-1}\rho_{ii}^\pm\right),
\end{eqnarray}
\begin{eqnarray}\label{ABBexplicit}
S^\pm(A\vert B)&=&-\rho_{11}^\pm\log\rho_{11}^\pm-\sum\limits_{j=2}^{L-1}\rho_{jj}^\pm\log\left(\sum\limits_{i=2}^{L-1}\rho_{ii}^\pm\right)\nonumber\\
&&+\sum\limits_{j=1}^{L-1}\rho_{jj}^\pm\log\left(\sum\limits_{i=1}^{L-1}\rho_{ii}^\pm\right).
\end{eqnarray}
Taking into account (\ref{ZBexplicit})-(\ref{ABBexplicit}) we see that (\ref{eq:BobsIgnorance}) in the single excitation cases reduces to the time-independent identity $1=1$.

\subsection{Quantum memory in two excitation basis}

We proceed with the two excitation case, and for the sake of an analytic approach, we consider the spin chain of $L=4$ length. We assume that the system initially is prepared in the product state 
$\rho_{ACDB}(0)=\ket{1100}\bra{1100}$ and after cumbersome calculations obtain the time evolved state 

\begin{eqnarray}\label{time evolved stateACDB}
&&\rho^\pm_{ACDB}(t)=\ket{\Phi^\pm(t)}\bra{\Phi^\pm(t)},\nonumber\\
&&\ket{\Phi^\pm(t)}=\sum\limits_{n=1}^5a^{\pm}_n \, \mathrm{e}^{\im t\Omega^{\pm}_n}\ket{\psi^{\pm}_n}.
\end{eqnarray}
Here we introduced notations: $\ket{\psi^\pm_1}=\alpha^\pm(\ket{1100}-\im \eta^\pm\ket{1010}
-\ket{1001}-\ket{0110}+\im \eta^\pm\ket{0101}+\ket{0011})$,
$\ket{\psi^\pm_2}=\gamma^\pm(\ket{1100}-\im \lambda^\pm\ket{1010}
-\ket{1001}-\ket{0110}+\im \lambda^\pm\ket{0101}+\ket{0011})$,
$\ket{\psi^\pm_3}=\frac{1}{\sqrt{6}}(\ket{1100}+\ket{1010}
+\ket{1001}+\ket{0110}+\ket{0101}+\ket{0011})$,
$\ket{\psi^\pm_4}=\frac{1}{\sqrt{12}}(\ket{1100}-2\ket{1010}
+\ket{1001}+\ket{0110}-2\ket{0101}+\ket{0011})$,
$\ket{\psi^\pm_5}=-\frac{1}{\sqrt{2}}\ket{1100}+\frac{1}{\sqrt{2}}\ket{0011}$,
$\textbf{a}^\pm=(\alpha^\pm, \gamma^\pm, 1/\sqrt{6}, 1/\sqrt{12}, -1/\sqrt{2})$, $|\textbf{a}^\pm|=
1$, where:
$\Omega^\pm_1=4J_2-2J_1-2\sqrt{(J_1-4J_2)^2+8D^2}$, 
$\Omega^\pm_2=4J_2-2J_1+2\sqrt{(J_1-4J_2)^2+8D^2}$,
$\Omega^\pm_3=4J_1+4J_2$,  $\Omega^\pm_4=-8J_1+4J_2$, $\Omega^\pm_5=-4J_2$, and extra notations read:
$\alpha^\pm=(4+2(\eta^{\pm})^2)^{-1/2}$, $\eta^\pm=(J_1-4J_2)/(\pm2D)-\sqrt{(J_1-4J_2)^2+8D^2}/(\pm2D)$, 
$\gamma^\pm=(4+2(\lambda^{\pm})^2)^{-1/2}$, $\lambda^\pm=(J_1-4J_2)/(\pm2D)+\sqrt{(J_1-4J_2)^2+8D^2}/(\pm2D)$. Here $\lambda^{\pm}\eta=-2$ and in what follows we consider $J_1=-1$, $J_2=1$, $D=1/2$. Tracing out parts $C$ and $D$, after laborious calculations we deduce reduced density matrix $\hat\varrho_{AB}^\pm(t)=\text{Tr}_{CD}\left[ \rho^\pm_{ACDB}(t)\right] $ which we present in terms of the Dirac matrices
\begin{eqnarray}\label{Dirac matrices}
&&\hat \varrho_{AB}^\pm=\frac{1}{4}\sum\limits_{\mu,\nu=0}^3\mathcal{R}_{\mu\nu}D_{\mu\nu},\nonumber\\
&&\mathcal{R}^\pm_{\mu\nu}=\Tr\left[\hat\varrho D_{\mu\nu}\right],~~\hat\sigma_0=\mathcal{I}_{A,B},\nonumber\\
&&D_{\mu\nu}=\hat\sigma^\mu_A\otimes\hat\sigma^\nu_B, 
\end{eqnarray}
where $D_{\mu\nu}$ is the Dirac matrix. In the explicit form:
%\begin{widetext}
\begin{eqnarray}\label{Rmatrix1}
 &&\mathbf{\mathcal{R}}=
\begin{bmatrix}
1 & 0 & 0 & \rho^{\pm}_{11} - \rho^{\pm}_{22} + \rho^{\pm}_{33} - \rho^{\pm}_{44}\\
0 &  2\text{Re}(\rho_{23}^{\pm}) & 2\text{Im}(\rho_{23}^{\pm}) & 0\\
0 & -2\text{Im}(\rho_{23}^{\pm}) & 2\text{Re}(\rho_{23}^{\pm}) & 0\\
\rho^{\pm}_{11} + \rho^{\pm}_{22} - \rho^{\pm}_{33} - \rho^{\pm}_{44} & 0 &0 & \rho^{\pm}_{11} - \rho^{\pm}_{22} - \rho^{\pm}_{33} + \rho^{\pm}_{44}
\end{bmatrix}.\nonumber
\end{eqnarray}
%\end{widetext}
\begin{align}\label{reducedAB} \nonumber
\hat \varrho_{AB}^\pm & = \rho^\pm_{11}(t)\ket{0}\bra{0}_A\otimes\ket{0}\bra{0}_B + \rho^\pm_{22}(t)\ket{1}\bra{1}_A\otimes\ket{0}\bra{0}_B\\
& +\rho^\pm_{33}(t)\ket{0}\bra{0}_A\otimes\ket{1}\bra{1}_B+\rho^\pm_{44}(t)\ket{1}\bra{1}_A\otimes\ket{1}\bra{1}_B\nonumber\\ 
& + \rho^\pm_{23}(t)\ket{1}\bra{0}_A\otimes\ket{0}\bra{1}_B+\rho^\pm_{32}(t)\ket{0}\bra{1}_A\otimes\ket{1}\bra{0}_B 
\end{align}
The time-dependent coefficients in (\ref{reducedAB}) are involved and presented in the appendix. 
Taking into account (\ref{Dirac matrices}) we calculate entropies:
\begin{eqnarray}\label{four spin case analytic expressions of entropies}
&& S^\pm(X\vert B)=S^{\pm}(\hat\varrho_{X,AB}^\pm)-S^{\pm}(\hat\varrho^\pm_{X,B}),\\
&& S^\pm(Z\vert B)=S^{\pm}(\hat\varrho_{Z,AB}^\pm)-S^{\pm}(\hat\varrho_{Z,B}^\pm),\\
&& S^\pm(A\vert B)=S^{\pm}(\hat\varrho_{AB}^\pm)-S^{\pm}(\hat\varrho_B^\pm),
\end{eqnarray}
\begin{figure}[t]
\centering
\includegraphics[width=0.7\linewidth]{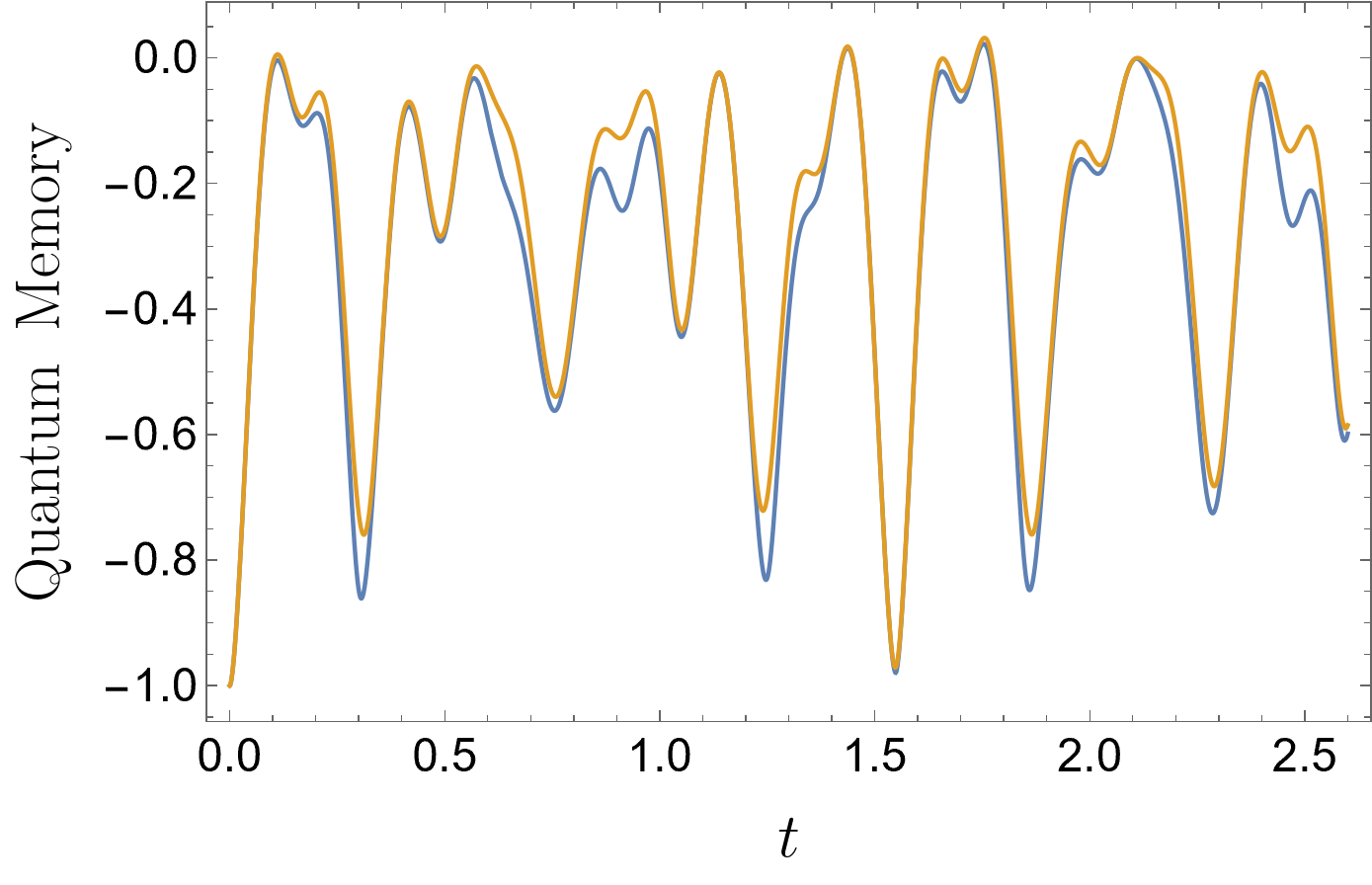}
\caption{Plot for the left hand side (amber) and the right hand side (blue), as functions of time for the  inequality $S^+(X|B) + S^+(Z|B)\geq \mathrm{log}_2(1/c) + S^+(A|B)$, (\ref{eq:BobsIgnorance}), where $c=1/2${} is a measure of complementarity and $S^+(X\vert B)$, $S^+(Z\vert B)$ are conditional quantum entropies of the states (\ref{eq:rhoZ}), (\ref{eq:rhoX}), $\ket{\psi_{1,2}}\equiv \ket{0}_A,\,\ket{1}_A$ and $\ket{\phi_{1,2}}=\frac{1}{\sqrt{2}}\left(\ket{0}_A\pm\ket{1}_A\right)$ respectively, the eigenfunctions of the $z$, $x$ components of the qubit $\textbf{A}$. The time unit is set by $|J_1|/\hbar$. In the time intervals when the left-hand side is larger than the right-hand side, quantum memory reduces the measurement uncertainty.}
\label{Entropyplot}
\end{figure}
in the explicit form:
\begin{eqnarray}\label{four spin case in the explicit form}
S^\pm(X\vert B)&=&-\sum\limits_{w^{X}_n\in[w^{X}_1,\ldots ,w^{X}_4]}w^{X}_n\log(w^{X}_n)+\nonumber\\
&& + \sum\limits_{y_n\in[y_1,y_2]}y_n\log(y_n),\\
S^\pm(Z\vert B)&=&-\sum\limits_{x^{Z}_n\in[x^{Z}_1,\ldots ,x^{Z}_4]}x^{Z}_n\log(x^{Z}_n)+\nonumber\\
&& + \sum\limits_{y_n\in[y_1,y_2]}y_n\log(y_n),\\
S^\pm(A\vert B) & = & -\sum\limits_{x_n\in[x_1,x_2,x_3,x_4]}x_n\log(x_n)+\nonumber\\
&& + \sum\limits_{y_n\in[y_1,y_2,y_3,y_4]}y_n\log(y_n).
\end{eqnarray}
Here we introduced the following notations
\begin{eqnarray}\label{four spins two excitation notations}
&&h(x)=-x\log (x)-(1-x)\log (1-x),\\ 
&& \nonumber \\
&&y_{1,2}=\rho^{\pm}_{11,(33)}(t)+\rho^{\pm}_{22,(44)}(t)\, ,\nonumber\\
&&x^Z_1=\rho^{\pm}_{11}(t);\,x^Z_{2}=\rho^{\pm}_{22}(t) \, , \nonumber\\ 
&&x^Z_3=\rho^{\pm}_{33}(t);\,x^Z_{4}=\rho^{\pm}_{44}(t) \, . \nonumber\\ 
&&x_1=\rho^{\pm}_{11}(t);\,x_{4}=\rho^{\pm}_{44}(t)=\rho^{\pm}_{11}(t),\nonumber\\ 
&&x_{2,3}=\frac{1}{2} \left( (\rho^{\pm}_{22} + \rho^{\pm}_{33}) \pm \sqrt{(\rho^{\pm}_{22} - \rho^{\pm}_{33})^2 + 4 |\rho_{23}^{\pm}|^2} \right),
\end{eqnarray}
and 
\begin{eqnarray}\label{second four spins two excitation notations}
&&w^{X}_{1,2}=\frac{1}{2}\left(1 + \sqrt{(\rho^{\pm}_{22} - \rho^{\pm}_{33})^2 + 4 |\rho_{23}^{\pm}|^2} \right),\nonumber\\ 
&&w^{X}_{3,4}=\frac{1}{2}\left(1 - \sqrt{(\rho^{\pm}_{22} - \rho^{\pm}_{33})^2 + 4 |\rho_{23}^{\pm}|^2} \right).
\end{eqnarray}

\noindent The results for EUR clockwise($+$)  and anticlockwise($-$) propagation cases, on two excitation basis, are plotted in Figure~\ref{Entropyplot}. As we see in the two excitation cases, EUR becomes time-dependent. The nonzero gap between  LHS and RHS that occurs at instant moments manifests reduced uncertainty of the quantum measurements and temporal character of the quantum memory. The result for the left($-$) propagating dispersion relation is similar and is not shown.

\subsection{OTOC in single excitation basis}

\begin{figure}[t]
    \centering
    \includegraphics[scale=0.7]{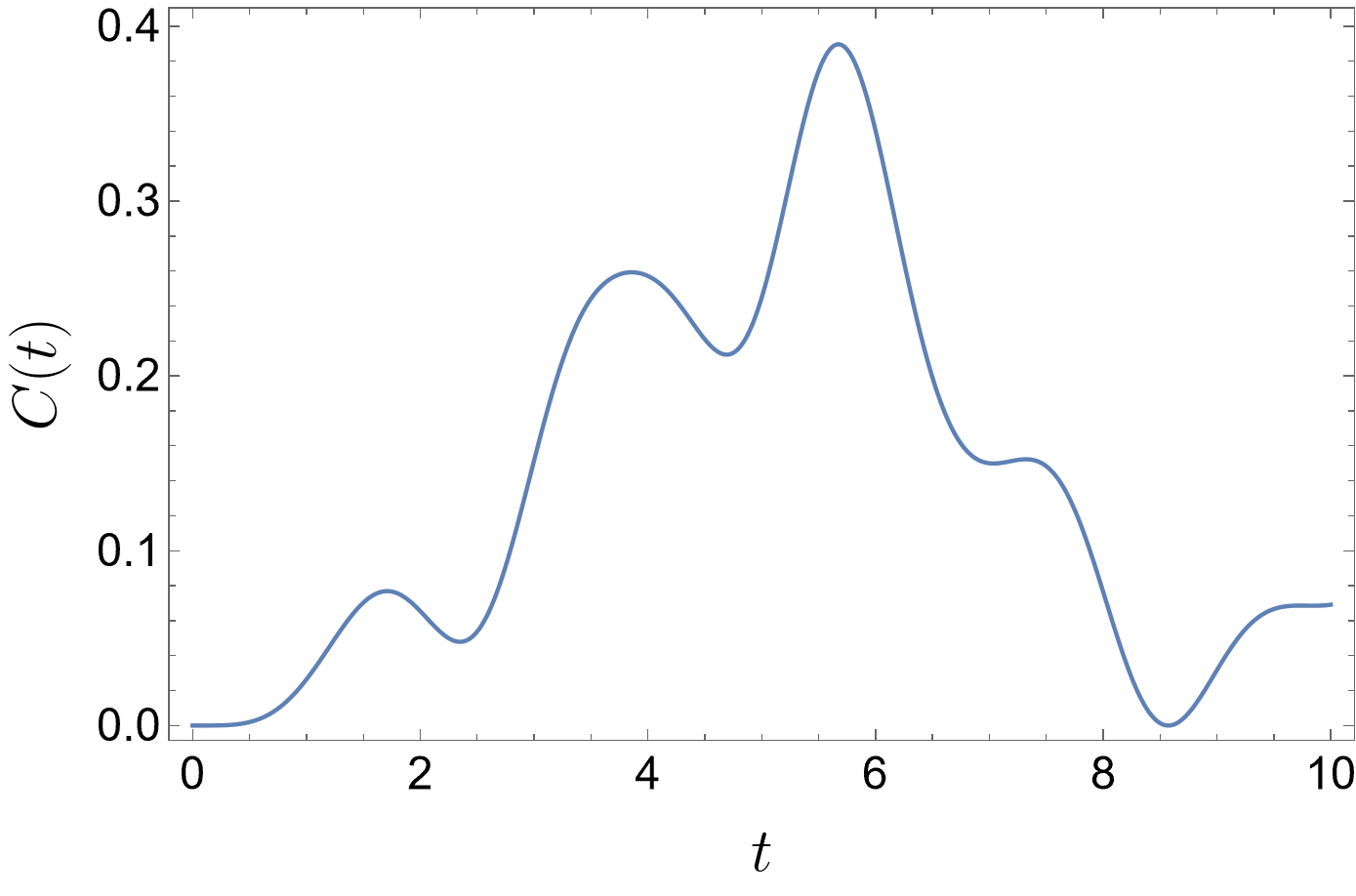} %\includegraphics[scale=0.58]{Fig5.png}
    \caption{Plot for $C^+(t)${}, in (\ref{wavefunctionone}) and the first excited state $n_1=1$, where we take $|\varphi_{n_i}\rangle \equiv n_i$. The spin chain has length, $L=10$ and perform summation over all states $n_{2},\, n_3 ,\, n_4$. For the coefficient $\mathcal{A}_{n_1n_2}^{n_3n_4}${}, use expressions in (\ref{1matrix elements})-(\ref{4matrix elements}), taking into account (\ref{eigenstates}).For phase we use (\ref{phase}). The time unit is set by $|J_1|/\hbar$.} \label{fig:OTOC}
\end{figure}
We assume that the system is prepared in the given eigenstate $\ket{n_1}=\ket{\varphi_{n_1}}$. Taking into account (\ref{eigenstates}) for matrix elements we deduce:
\begin{figure}[h]
     \centering
     \begin{subfigure}[b]{0.38\linewidth}
         \centering
         \includegraphics[width=\linewidth]{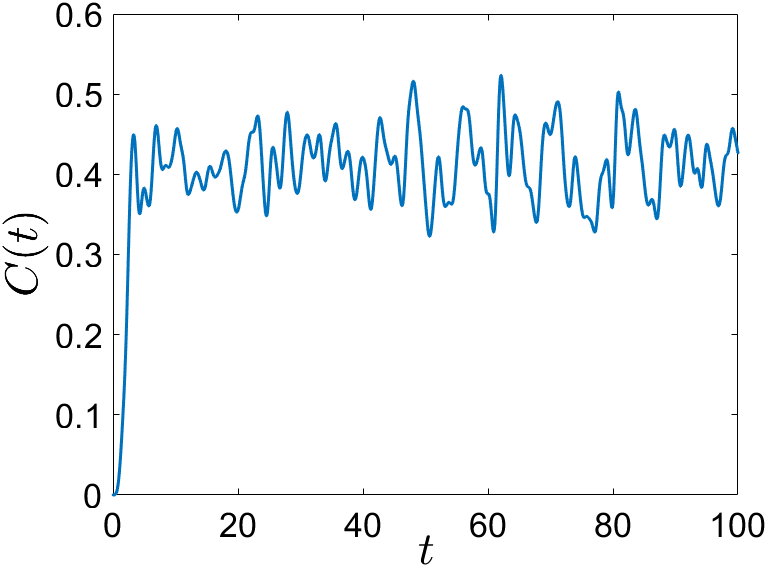}
         \caption{$L=10$}
         \label{fig:fig6a}
     \end{subfigure}
     %\hfill
	    \begin{subfigure}[b]{0.38\linewidth}
         \centering
         \includegraphics[width=\linewidth]{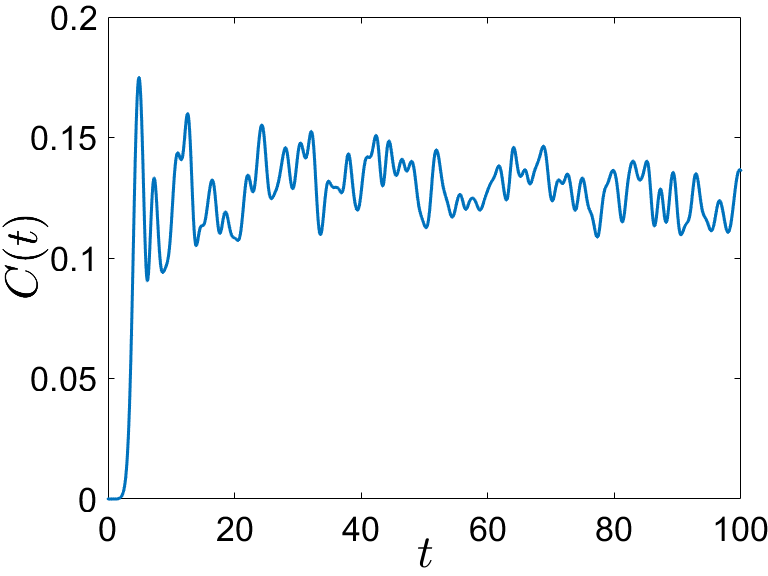}
         \caption{$L=20$}
         \label{fig:fig6b}
     \end{subfigure}
     \begin{subfigure}[b]{0.38\linewidth}
         \centering
         \includegraphics[width=\linewidth]{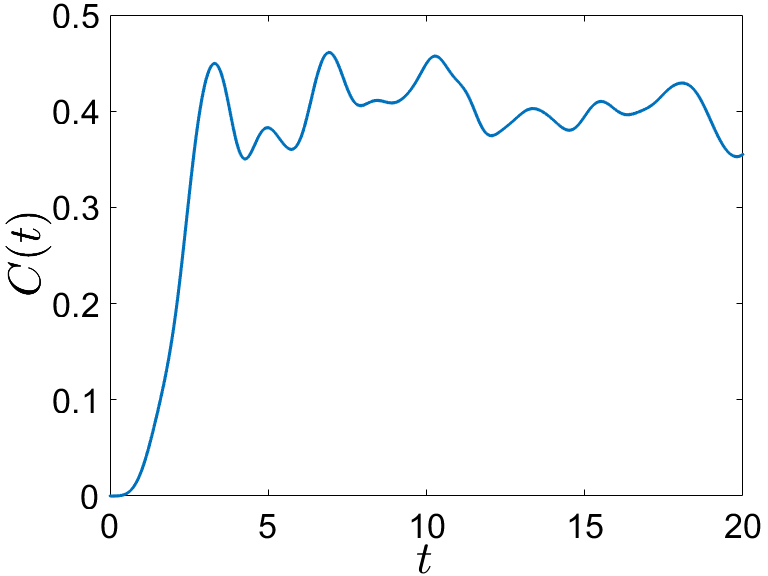}
         \caption{$L=10$}
         \label{fig:fig6c}
     \end{subfigure}
     %\hfill
	    \begin{subfigure}[b]{0.38\linewidth}
         \centering
         \includegraphics[width=\linewidth]{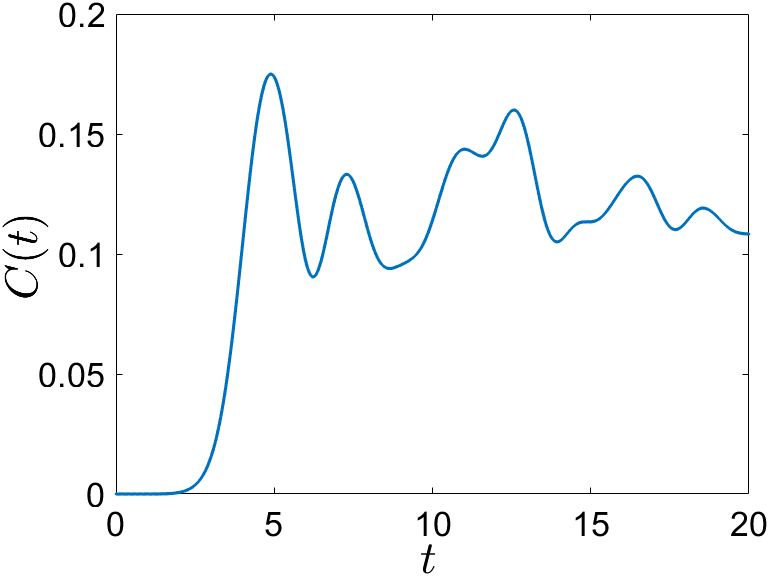}
         \caption{$L=20$}
         \label{fig:fig6d}
     \end{subfigure}
     \caption{OTOC as a function of time for two excitation case. The number of the spins $L=10$ (\ref{fig:fig6a}) and $L=20$ (\ref{fig:fig6b}).  Figure \ref{fig:fig6c} and Figure \ref{fig:fig6d} are the same but plotted for a shorter time interval.  The system parameters are $J_1 = -1$, $J_2 = 1$, $D=0.5$. The time unit is set by $|J_1|/\hbar$.}
     \label{OTOCtwoJ1J2}
\end{figure}

\begin{eqnarray}\label{1matrix elements}
&&\bra{n_1}\hat\sigma_{L/2}^z\ket{n_2}=\frac{1}{L}\sum\limits_{j=1}^L \mathrm{e}^{\im 2\pi(n_1-n_2)j/L}\left(1-2\delta_{L/2,j}\right),
\end{eqnarray}
\begin{eqnarray}\label{2matrix elements}
&&\bra{n_2}\hat\sigma_{1}^z\ket{n_3}=
\frac{1}{L}\sum\limits_{j=1}^L \mathrm{e}^{\im 2\pi(n_2-n_3)j/L}\left(1-2\delta_{1,j}\right),
\end{eqnarray}
\begin{eqnarray}\label{3matrix elements}
&&\bra{n_3}\hat\sigma_{L/2}^z\ket{n_4}=\frac{1}{L}\sum\limits_{j=1}^L \mathrm{e}^{\im 2\pi(n_3-n_4)j/L}\left(1-2\delta_{L/2,j}\right),
\end{eqnarray}
\begin{eqnarray}\label{4matrix elements}
&&\bra{n_4}\hat\sigma_{1}^z\ket{n_1}=\frac{1}{L}\sum\limits_{j=1}^L \mathrm{e}^{\im 2\pi(n_4-n_1)j/L}\left(1-2\delta_{1,j}\right),
\end{eqnarray}
and 
\begin{eqnarray}\label{phase}
&&E^\pm_{n_1}-E^\pm_{n_2}+E^\pm_{n_3}-E^\pm_{n_4}=\nonumber\\
&&J_1\cos(2\pi n_1/L)+J_2\cos(4\pi n_1/L)\pm D\sin(2\pi n_1/L)\nonumber\\
&&-J_1\cos(2\pi n_2/L)-J_2\cos(4\pi n_2/L)\mp D\sin(2\pi n_2/L)\nonumber\\
&&+J_1\cos(2\pi n_3/L)+J_2\cos(4\pi n_3/L)\pm D\sin(2\pi n_3/L)\nonumber\\
&&-J_1\cos(2\pi n_4/L)-J_2\cos(4\pi n_4/L)\\ 
&& \pm D\sin(2\pi n_4/L).\nonumber
\end{eqnarray}
Comparing EUR calculated for two excitation basis Figure \ref{Entropyplot} with the OTOC calculated for the single excitation case Figure \ref{fig:OTOC} we see that EUR oscillates in time much faster. The left propagating OTOC $C^{-}(t)$ shows a slight shift in time (not shown). We note that in a single excitation case, OTOC does not thermalize to a steady state, and the large amplitude oscillations last for an arbitrarily long time.

\subsection{OTOC in two excitation basis}

\begin{figure}[t]
     \centering
     \begin{subfigure}[b]{0.38\linewidth}
         \centering
         \includegraphics[width=\linewidth]{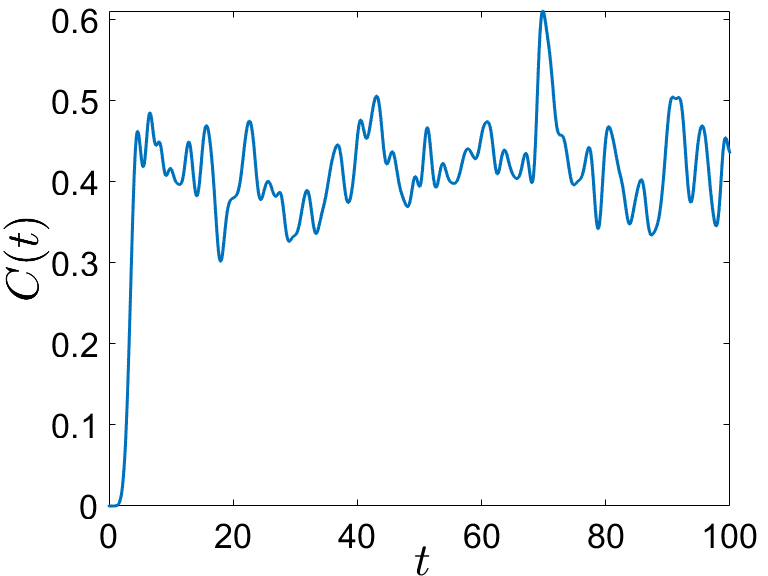}
         \caption{$L=10$}
         \label{fig:figa}
     \end{subfigure}
     %\hfill
	    \begin{subfigure}[b]{0.38\linewidth}
         \centering
         \includegraphics[width=\linewidth]{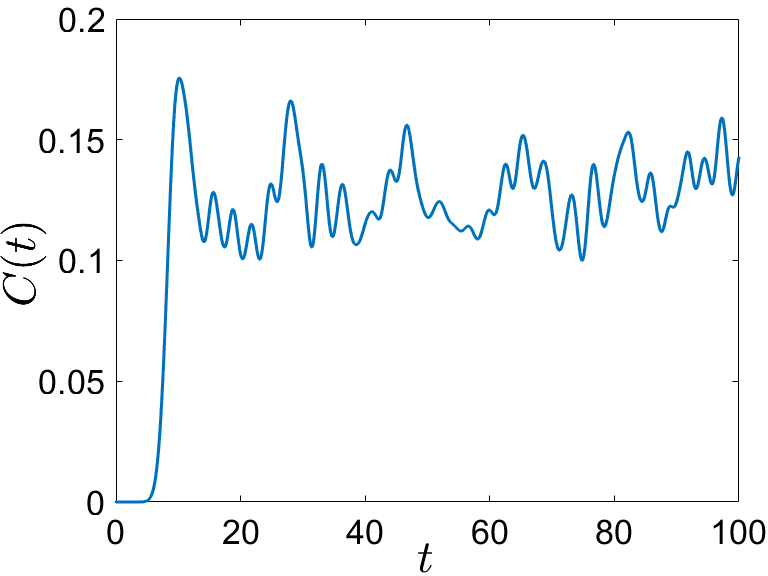}
         \caption{$L=20$}
         \label{fig:figb}
     \end{subfigure}
     \begin{subfigure}[b]{0.38\linewidth}
         \centering
         \includegraphics[width=\linewidth]{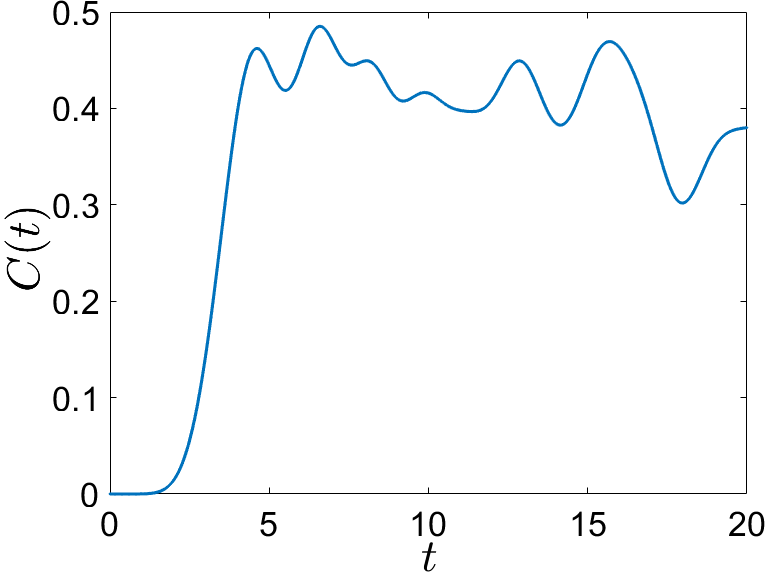}
         \caption{$L=10$}
         \label{fig:figc}
     \end{subfigure}
     %\hfill
	    \begin{subfigure}[b]{0.38\linewidth}
         \centering
         \includegraphics[width=\linewidth]{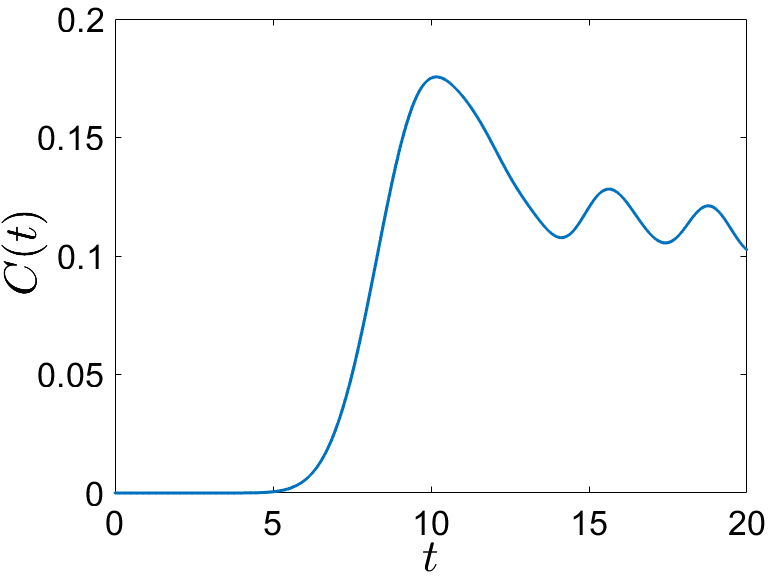}
         \caption{$L=20$}
         \label{fig:figd}
     \end{subfigure}
     \caption{OTOC as a function of time for two excitation case and nearest neighbor coupling. The number of the spins $L=10$ (\ref{fig:figa}) and $L=20$ (\ref{fig:figb}).  Figures \ref{fig:figc}, \ref{fig:figd} are the same but plotted for a shorter time interval.  The system parameters are $J_1 = -1$, $J_2 = 0$, $D=0.5$. The time unit is set by $|J_1|/\hbar$.}
%     \caption{OTOC as a function of time for two excitation case and nearest neighbor coupling. The number of the spins $L=10$ (top,left) and $L=20$ (top,right).  Bottom left/right are the same but plotted for a shorter time interval.  The system parameters are $J_1 = -1$, $J_2 = 0$, $D=0.5$. The time unit is set by $|J_1|/\hbar$.}
\label{OTOConeJ1J2}
\end{figure}

The results for OTOC $C^{+}(t)$ in the case of two excitations in the system are plotted in Figure \ref{OTOCtwoJ1J2}. Again, we see that OTOC shows slower time oscillations than quantum memory, Figure \ref{Entropyplot}. However, as compared to the single excitation case, we see the relaxation in the system after $t>5$. The result for $C^{-}(t)$ is the same except for a slight delay in time.  We explore OTOC in the two excitation case for the next nearest coupling $J_2=0$. The system (\ref{Hamiltonian2}) reduces to the $XXZ$ model with the anisotropy parameter $\Delta=\sqrt{J_1^2+D^2}/J_1$. Following \cite{essler2005one}, we present the solution to the two-excitation problem in the following form: 
\begin{eqnarray}\label{Bethe1}
\ket{\Psi_n}=\sum\limits_{j,j'\neq j}^Lf[j,j']\ket{j,j'},
\end{eqnarray}
where $\ket{j,j'}$ means excitation on sites $j$, $j'$ and otherwise zeros, amplitudes have the form:
\begin{eqnarray}\label{Bethe2}
&&f[j,j']=\exp[\im k_1j+\im k_2j'+\theta(k_1,k_2)]+\nonumber\\
&&\exp[\im k_2j+\im k_1j'-\theta(k_1,k_2)],\\
&& \nonumber\\
&&\theta(k_1,k_2)=2\arctan\left[\frac{\Delta\sin(k_1-k_2)/2}{\cos(k_1+k_2/2)-\Delta\cos(k_1-k_2)/2}\right],\nonumber\\
&&k_1=(2\pi/N)I_1-\theta(k_1,k_2)/N,\,\,\,\,\,I_1\in [1,N-1],\nonumber\\
&&k_2=(2\pi/N)I_2-\theta(k_1,k_2)/N,\,\,\,\,\,I_2\in [1,N-1],
\end{eqnarray}\\

\noindent and quantum number $n$ is given by the pair $n=(I_1,I_2)$. 
Taking into account (\ref{Bethe1}) we calculate the matrix elements in (\ref{wavefunctionone}). The result for OTOC $C^{+}(t)$ is plotted in Figure \ref{OTOConeJ1J2}. Compared to the  $J_2\neq 0$ case, in Figure \ref{OTOCtwoJ1J2},  the oscillation amplitude is slightly larger in the stationary case $t>5$. 
%\begin{figure*}
%    \centering
%\includegraphics[width=0.45\linewidth,height=0.30\linewidth]{misra_otoc_figs/C_2_N10_J2_0.png}
%\includegraphics[width=0.45\linewidth,height=0.30\linewidth]{misra_otoc_figs/C_2_N20_J2_0.png}
%\includegraphics[width=0.45\linewidth,height=0.30\linewidth]{misra_otoc_figs/C_2_N10_J2_0_sort.png}
%\includegraphics[width=0.45\linewidth,height=0.30\linewidth]{misra_otoc_figs/C_2_N20_J2_0_sort.png}
%    \caption{OTOC as a function of time for two excitation case and nearest neighbor coupling. The number of the spins $L=10$ (top,left) and $L=20$ (top,right).  Bottom left/right are the same but plotted for a shorter time interval.  The system parameters are $J_1 = -1$, $J_2 = 0$, $D=0.5$. The time unit is set by $|J_1|/\hbar$.}
%    \label{}
%\end{figure*}

\subsection{Neural network predictions for OTOC and quantum memory}

%\begin{figure}[h]
%\centering
%\includegraphics[width=0.3\linewidth]{Fig6.png}
%\caption{The pictorial plot of the artificial neurons which are used to predict the time-dependence of OTOC and quantum memory.}
%\end{figure}
We trained the artificial neural network to predict results for OTOC and quantum memory for the left and right propagating spin waves. The set of parameters was defined as follows: the length of the spin chain $L$, the ferromagnetic and antiferromagnetic exchange constants $J_1$ and $J_2$, respectively, and the DMI interaction constant $D$, and the excitation number $n_1=1$. Predictions were made over the first 10 time units with a resolution of 0.4 time steps, resulting in 25 predicted points per curve. The data set consisted of 1331 OTOC curves, generated by varying the parameters within the following ranges: $J_1 \in (-1, 0), J_2 \in (0, 1)$, and $\Delta \in (0, 0.5)$. About 80 percent of the data was used for training, while the remaining 20 percent was reserved for testing the model's predictive performance. The neural network architecture comprised two hidden \textit{fully connected layers} with 64 and 32 neurons, respectively, both using the \textit{ReLU activation function}. The output layer consisted of 25 neurons, each corresponding to a time step in the predicted OTOC curve (see Figure~\ref{ANN_architecture}).
\begin{figure}[t]
     \centering
     \begin{subfigure}[b]{0.48\linewidth}
         \centering
         \includegraphics[width=\linewidth]{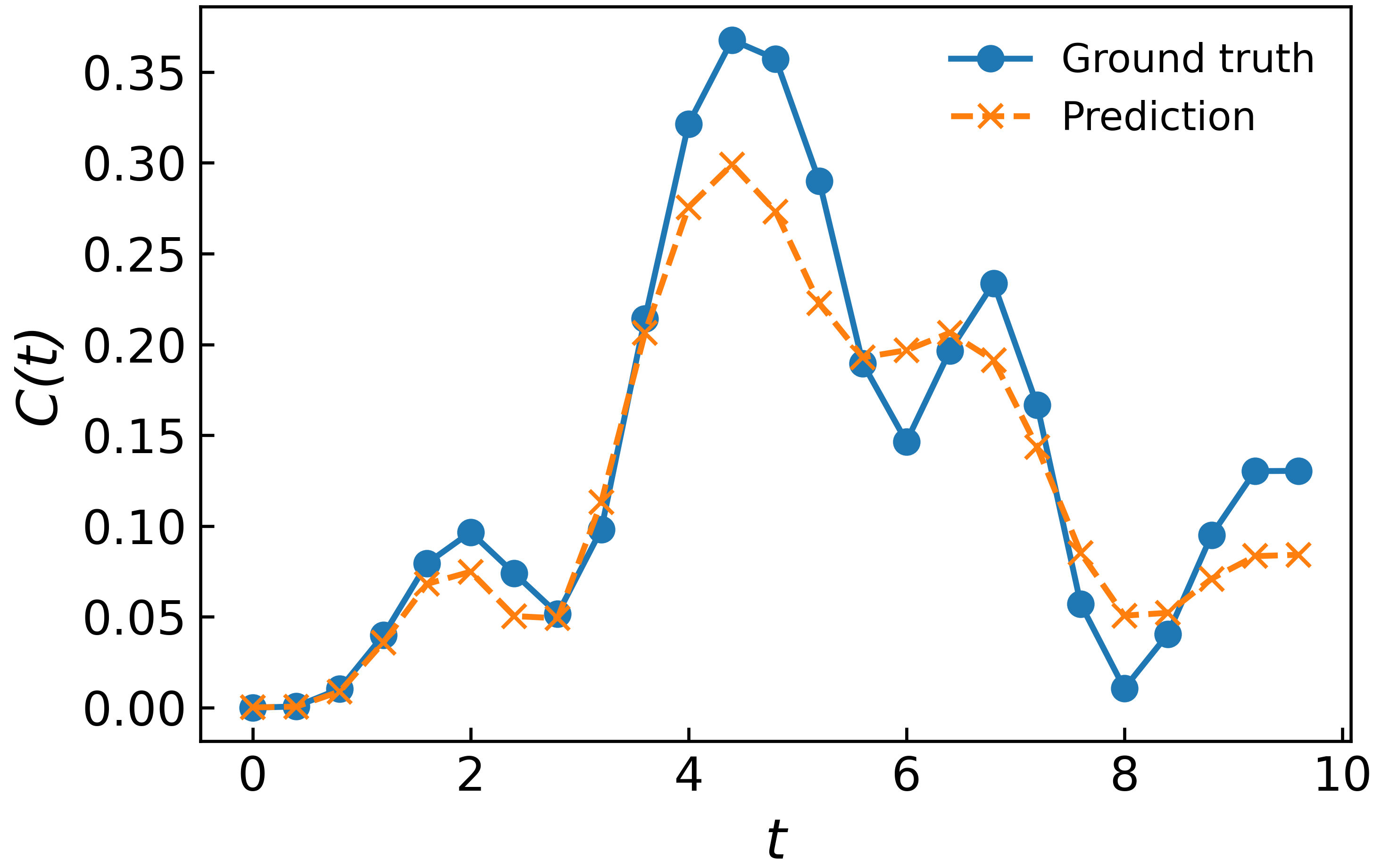}
         \caption{The exact and predicted OTOC curves for exchange constants $J_1 = -0.70$, $J_2 = 1.00$, and DMI $D =0.30$.}
         \label{OTOC_pred_1}
     \end{subfigure}
     \hfill
	    \begin{subfigure}[b]{0.48\linewidth}
         \centering
         \includegraphics[width=\linewidth]{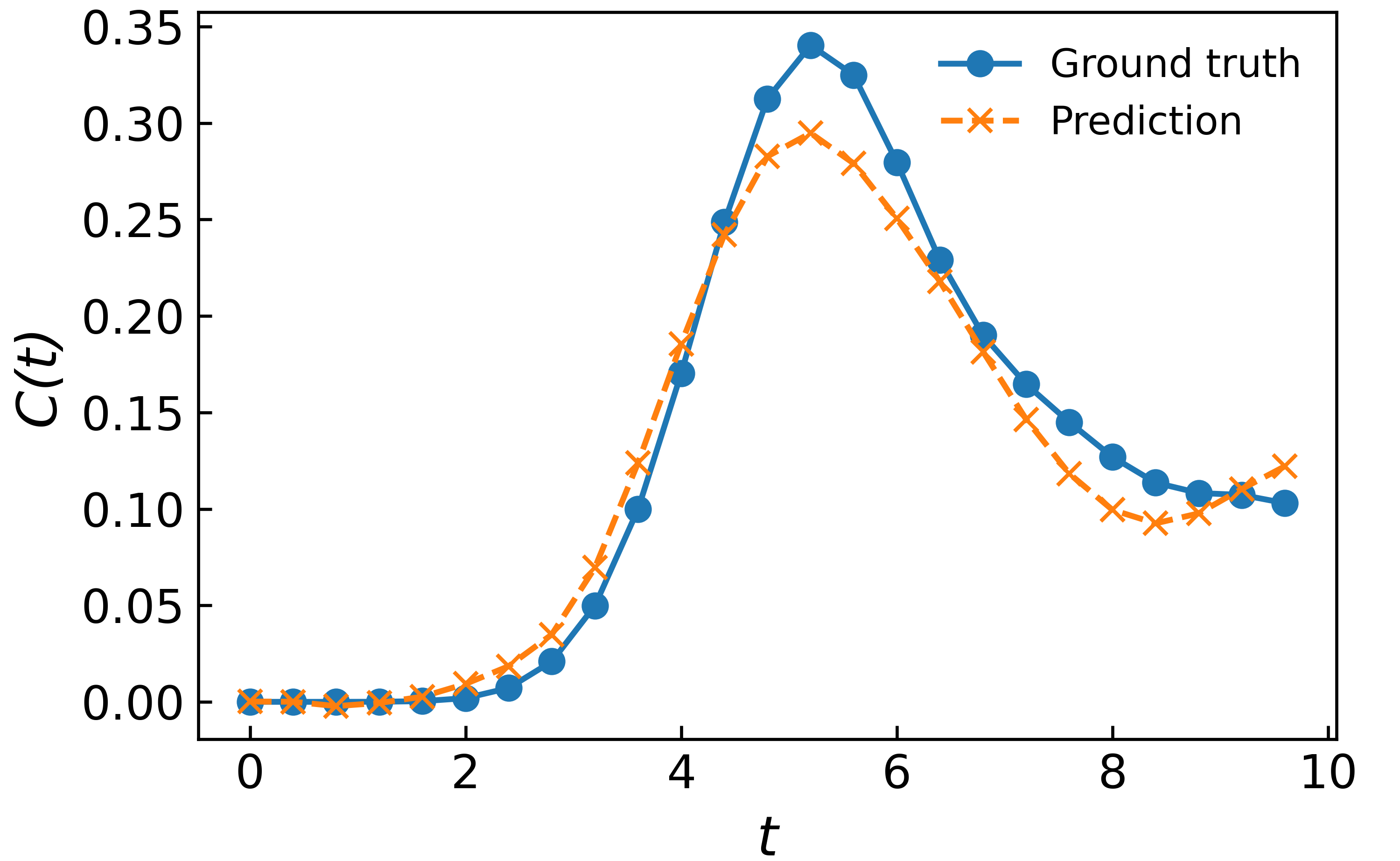}
         \caption{The exact and predicted OTOC curves for exchange constants $J_1 = -0.90$, $J_2 = 0.00$, and DMI $D =0.40$.}
         \label{OTOC_pred_2}
     \end{subfigure}
     \begin{subfigure}[b]{0.48\linewidth}
         \centering
         \includegraphics[width=\linewidth]{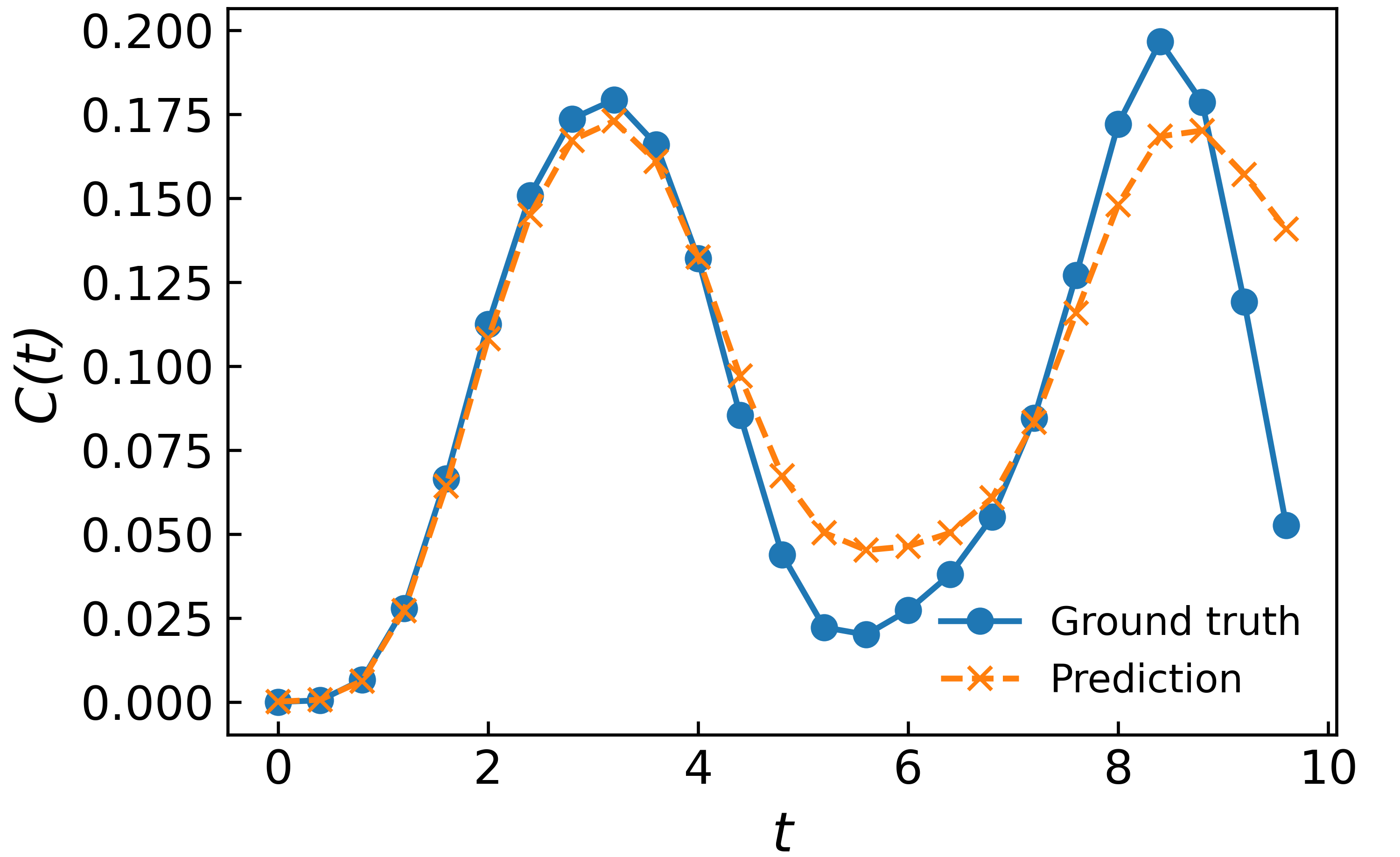}
         \caption{The exact and predicted OTOC curves for exchange constants $J_1 = -0.20$, $J_2 = 0.90$, and DMI $D =0.35$.}
         \label{OTOC_pred_3}
     \end{subfigure}
     \hfill
	    \begin{subfigure}[b]{0.48\linewidth}
         \centering
         \includegraphics[width=\linewidth]{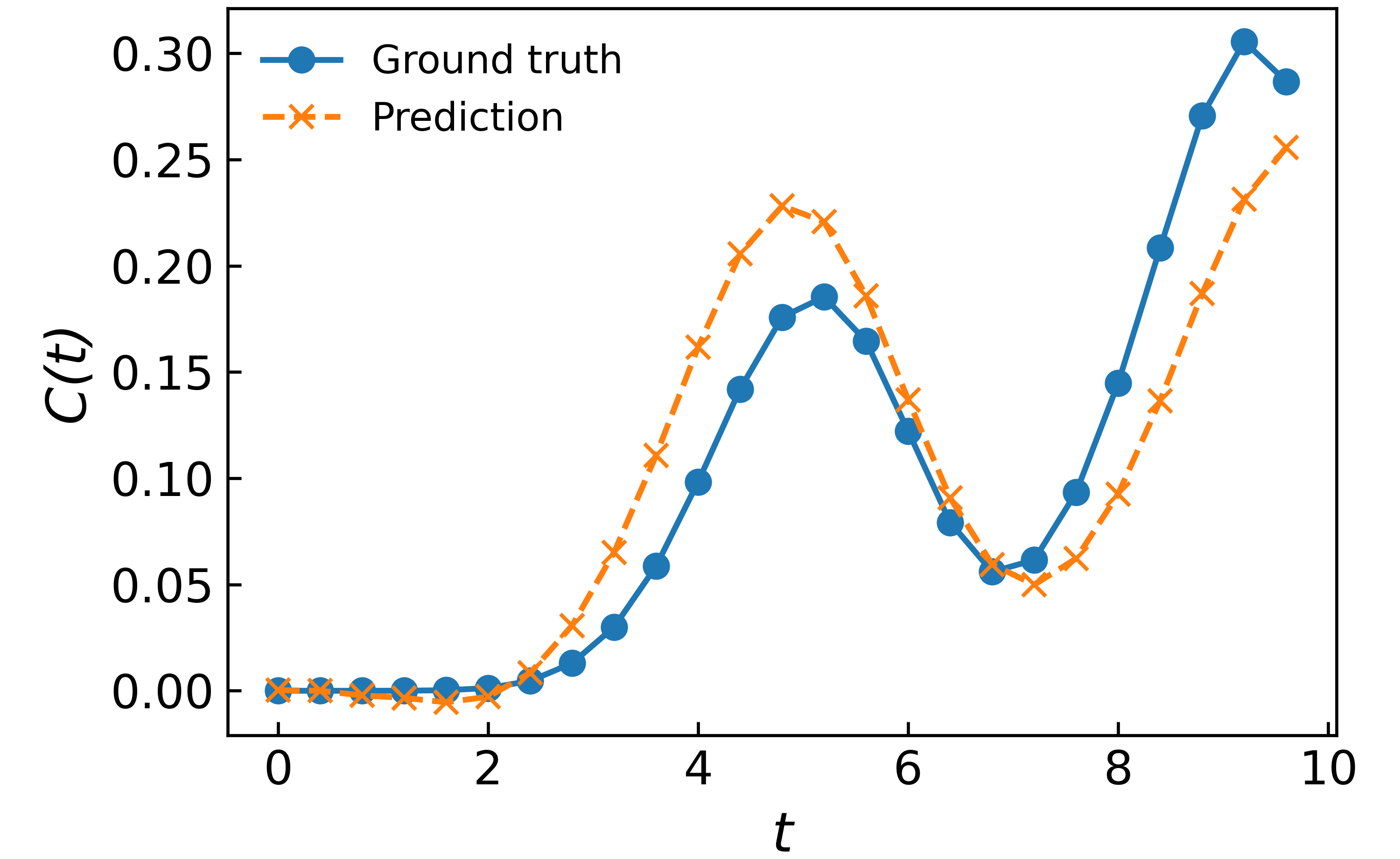}
         \caption{The exact and predicted OTOC curves for exchange constants $J_1 = -0.80$, $J_2 = 0.00$, and DMI $D =0.505$.}
         \label{OTOC_pred_4}
     \end{subfigure}
     \caption{The exact (ground truth) and the predicted via neural network values of OTOC as a function of time. For particular values of parameters specified in the captions of corresponding figures, respectively. As we see, the maximal values of the exact and predicted OTOCs are synchronized in time and show good qualitative agreement for all sets of parameters. }
     \label{figure8}
\end{figure}
The results of a prediction for OTOC, for selected values of parameters $J_{1,2}$ and $D$ are shown in Figures \ref{OTOC_pred_1}, \ref{OTOC_pred_2}, \ref{OTOC_pred_3}, \ref{OTOC_pred_4}. The goal of the artificial neural network was to learn and predict OTOC and quantum memory curves based on five parameters: the length of the spin chain $L=10$, ferromagnetic and antiferromagnetic exchange constants $J_1$, $J_2$, constant of the DM interaction $D$, and magnon excitation sector $n_1=1$. \vspace{0.2cm}\\
To mitigate overfitting \cite{Hinton_Dropout}, a dropout rate of 0.1 was applied after the first and second hidden layers. The predicted OTOC curves are shown in Figures \ref{OTOC_pred_1}, \ref{OTOC_pred_2}, \ref{OTOC_pred_3}, and \ref{OTOC_pred_4}. The same approach was implemented to predict LHS and RHS curves of the quantum memory. Figure \ref{OTOC_pred_1} and Figure \ref{OTOC_pred_2} present results for OTOC corresponding to the exact results (ground truth) and predicted by neural network results, respectively. Due to the selected set of parameters, the ground state in Figure \ref{OTOC_pred_1} is the helix and in 
Figure \ref{OTOC_pred_2} shows that the ground state is the slightly distorted ferromagnetic state (i.e., antiferromagnetic term leading to the spin frustration is zero $J_2=0$). In  Figure \ref{OTOC_pred_3}, the dominant term is antiferromagnetic exchange $J_2$, and in Figure \ref{OTOC_pred_4}, two terms, ferromagnetic exchange $J_1$ and strong DM interaction $D=0.5$, compete with each other, leading to the helical ground state. In all cases predicted by the neural network, values of OTOC are pretty close to the exact ground values, and peaks in time dependence are synchronized and are in phase (i.e., maximum and minimum values of the predicted and exact OTOCs coincide in time). 
\begin{figure}[t]
\centering
\includegraphics[width=0.68\linewidth]{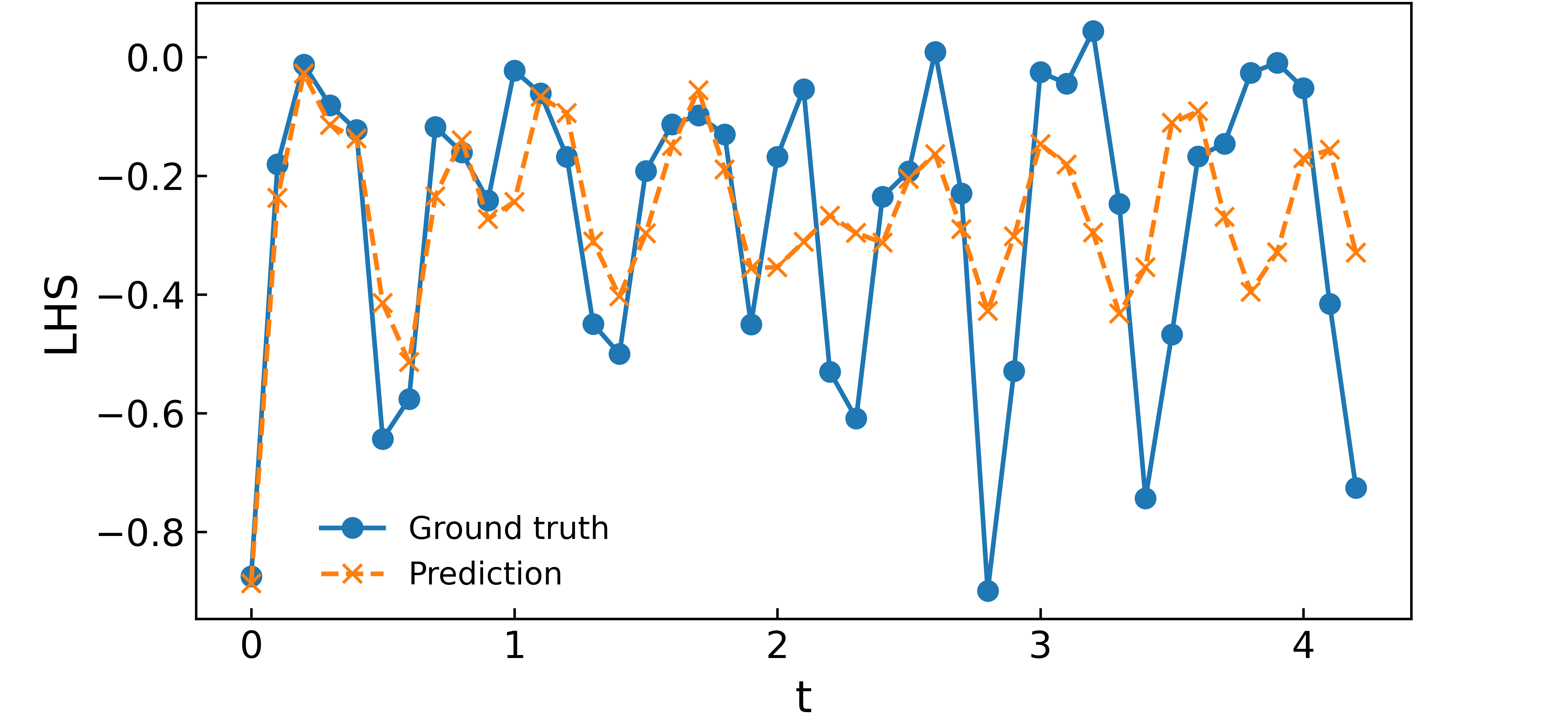}
\caption{The time-dependent quantum memory for anticlockwise (LHS) propagating spin excitations. The values of the parameters: $J_1 = -0.90, J_2 = 0.90, D =0.50$. We see growth in time mismatch between exact and predicted by neural network peaks of the quantum memory. The result for clockwise (RHS) propagating spin excitations manifests the same trend (not shown).}
\label{Mem_left_1}
\end{figure}
\begin{figure}[ht]
\centering
\includegraphics[width=0.68\linewidth]{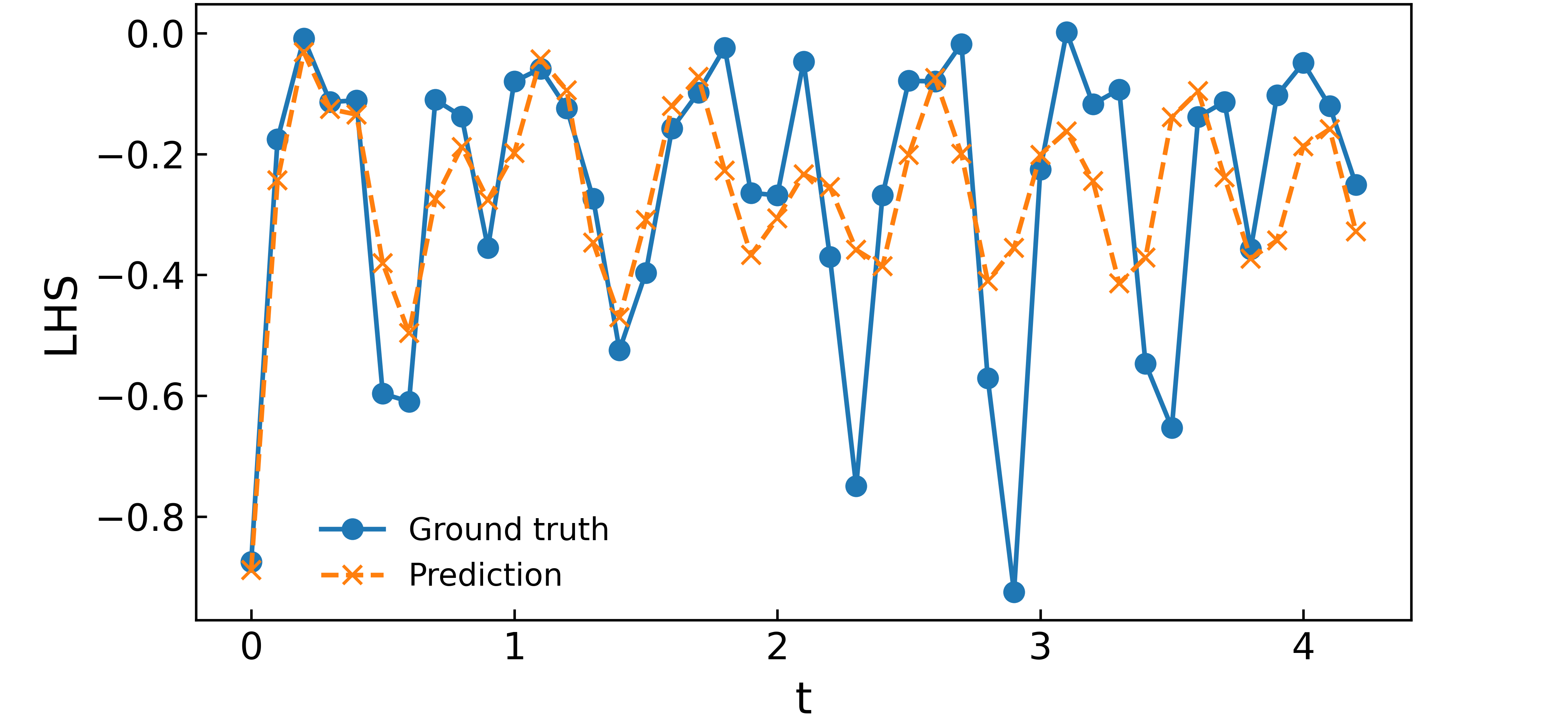}
\caption{The time-dependent quantum memory for anticlockwise (LHS) propagating spin excitations. The values of the parameters: $J_1 = -0.90, J_2 = 0.90, D =0.0$. When DM interaction is zero $D=0$, the mismatch between the exact and predicted via the neural network peaks and phases of the quantum memory signal are smaller than in the nonzero DM case. The result for clockwise (RHS) propagating spin excitations manifests the same trend (not shown).} 
\label{Mem_left_2}
\end{figure}
When it concerns the quantum memory Figures \ref{Mem_left_1}, \ref{Mem_left_2}, we see the phase shift between the exact and predicted values of quantum memory. The phase shift in the time dependence is accumulated in time for the large DM case $D=0.5$. The obtained result shows that the quantum memory is more sensitive than OTOC with respect to the broken inversion symmetry, nonreciprocal effect, and strong DM interaction.

\section{Coffman–Kundu–Wootters (CKW) monogamy inequality}

As we observed in previous sections, ANN predictions of Quantum Memory are quite sensitive to the DMI constant. In this section, we study the dependence of different physical characteristics, such as multipartite entanglement and chirality, on the DMI constant. In particular, for the measure of multipartite entanglement, we consider Coffman–Kundu–Wootters (CKW) monogamy inequality. We again consider a four-spin-$\tfrac12$ system with Hilbert space $\mathcal{H} = (\mathbb{C}^2)^{\otimes 4}$ and particular pure states $\hat\rho_{\ket{\psi}} = \ket{\psi_2}\bra{\psi_2}$, where the state $\ket{\psi_2}$ is defined in \textbf{section 3.2}, and we also consider another eigenstate: $\hat\rho_{\ket{\phi}} = \ket{\phi_2}\bra{\phi_2}$,  $\ket{\phi}=\frac{i}{2}\ket{1000}-\frac{1}{2}\ket{0100}-\frac{i}{2}\ket{0010}+\frac{1}{2}\ket{0001}$. As we mentioned in the introduction, the $z$ component of the vector chirality $\hat{\varkappa}^{z} = \hat{\varkappa}_{i}^{z} = (\mathbf{\hat S}_i \times \mathbf{\hat S}_{i+1})_z$, is the order parameter of the system. The expectation value of the order parameter can be calculated directly for the selected states and read $\bra{\phi}\hat{\varkappa}^{z}\ket{\phi}=1/4$, $\bra{\psi}\hat{\varkappa}^{z}\ket{\psi}=8\lambda^\pm\left(\gamma^\pm\right)^2$, where $\gamma^\pm=(4+2(\lambda^{\pm})^2)^{-1/2}$, $\lambda^\pm=(J_1-4J_2)/(\pm2D)+\sqrt{(J_1-4J_2)^2+8D^2}/(\pm2D)$. Coffman–Kundu–Wootters (CKW) monogamy inequality for pure states is defined as follows  \cite{Azimi2014,Coffman2000}: 
\begin{eqnarray}\label{monogamy inequality}
\tau_{i|rest}\ge\tau_{ij}+\tau_{ik}+\tau_{i\ell}.
\end{eqnarray}
Here 
\begin{eqnarray}\label{One-Tangle}
\tau_{i|rest} = 4\,\det(\rho_i),
\end{eqnarray}
is one-Tangle of the $i$ spin and two-Tangle between spins $i$ and $j$ is defined via the pairwise concurrence $C[\rho_{ij}]$:
\begin{eqnarray}\label{Two-Tangle}
\tau_{ij} = C^2[\rho_{ij}].
\end{eqnarray}
Concurrence we calculate using the standard protocol 
\begin{eqnarray}\label{1Concurrence}
C(\rho_{ij})=\max\left(0,R_1 - R_2 - R3 - R_4\right),
\end{eqnarray}
where $R_i$ are eigenvalues of the matrix $R = \rho_{ij}\tilde{\rho}_{ij}$, $\tilde{\rho}_{ij}=
(\sigma_y \otimes \sigma_y)\,\rho_{ij}^*\,(\sigma_y \otimes \sigma_y)$, in the decreasing order. Calculations done for the state $\hat\rho_{\ket{\phi}}$ we obtain: $\tau_{i|rest}=\tau_{ij}+\tau_{ik}+\tau_{i\ell}$, while for the state 
$\hat\rho_{\ket{\psi}}$ we obtain:
\begin{eqnarray}\label{2monogamy inequality}
\left(2-\lambda^2\right)^2\tau_{i|rest}=\left(2+\lambda^2\right)^2\left(\tau_{ij}+\tau_{ik}+\tau_{i\ell}\right).
\end{eqnarray}
We note that the state $\hat\rho_{\ket{\phi}}$ with zero multipartite entanglement does not depend on the DMI constant, while $\hat\rho_{\ket{\psi}}$ does. Consequently, for $\hat\rho_{\ket{\phi}}$, the total entanglement in the system $\tau_{i|rest}$ entirely factorizes into pairwise quantum entanglement, and the multipartite entanglement is zero. Whereas for $\hat\rho_{\ket{\psi}}$ and any nonzero $\lambda$, the total entanglement in the system $\tau_{i|rest}$ is larger than pairwise quantum entanglement, meaning that multipartite entanglement is not zero. Thus, not only the ANN predictions of the Quantum Memory, but also the system's multipartite entanglement and order parameter depend on the value of the DMI constant.

\section{Conclusions}

It is well known that \textit{Entropic uncertainty relations} are universal measures of fundamental uncertainties of quantum measurements and are widely discussed in the quantum metrology literature. In the seminal work  \cite{berta2010uncertainty}, authors introduced the concept of quantum memory, in essence, a specific type of quantum correlation that allows for the reduction of fundamental uncertainties of quantum measurements. The formalism of quantum-assisted EUR was developed for stationary systems only, and here, in the present work, we generalized this formalism for time-independent problems. We studied the emergence and propagation of quantum correlations in time in the helical spin system. We considered the atomic chain of spins coupled by the nearest-neighbor ferromagnetic and the next-nearest-neighbor antiferromagnetic interactions deposited on the Ir(001) surface. Competing exchange interaction terms lead to spin frustration, while the Dzyaloshinskii-Moriya (DM) interaction between nearest spins stabilizes the chiral spin order. We considered two measures to explore the buildup of quantum correlations in time. The first of those measures is the OTOC, widely discussed in the literature. The second measure is quantum memory, which we modified for time-dependent problems. We considered single and two excitation sectors in the system and observed the following differences in the time behavior of OTOC and quantum memory. Quantum memory manifests faster oscillations in time than OTOC. We implemented an artificial neural network to predict the values of quantum memory and OTOC. The results obtained via the artificial neural network suggest that quantum memory is more sensitive than OTOC regarding broken inversion symmetry, nonreciprocal effect, and strong DM interaction. Comparing the predicted via neural network values of quantum memory with its exact values, we found that the network's predictions are less accurate when DM interaction is stronger.

\appendix

\section{Density matrix elements for $\hat{\varrho}_{AB}(t)$}\label{appendixA}
The elements of the reduced density matrix in two excitation case, $\rho_{AB}(t)$ are:
\begin{eqnarray}\nonumber
 \rho^{\pm}_{11}  &=&  (\alpha^{\pm})^4+(\gamma^{\pm})^4+2 (\alpha^{\pm})^2 (\gamma^{\pm})^2 \cos [( \Omega_1-\Omega_2) t]\\ \nonumber
&& - \frac{1}{3}(\alpha^{\pm})^2 \cos [ (\Omega_1-\Omega_3) t]-\frac{1}{6} (\alpha^{\pm})^2 \cos [ (\Omega_1-\Omega_4) t]\\ \nonumber
&& -\frac{1}{3}(\gamma^{\pm})^2 \cos [ (\Omega_2-\Omega_3) t]-\frac{1}{6} (\gamma^{\pm})^2 \cos [(\Omega_2-\Omega_4) t]\\ 
&& +\frac{1}{36}\cos [(\Omega_3-\Omega_4) t]+\frac{5}{144}, 
\end{eqnarray}
\begin{eqnarray}\nonumber
 \rho^{\pm}_{44}  & = & (\alpha^{\pm})^4+(\gamma^{\pm})^4+2 (\alpha^{\pm})^2(\gamma^{\pm})^2 \cos [ \Omega_1-\Omega_2) t]\\ \nonumber
&& - \frac{1}{3}(\alpha^{\pm})^2 \cos [ (\Omega_1-\Omega_3) t]-\frac{1}{6} (\alpha^{\pm})^2 \cos [ (\Omega_1-\Omega_4) t]\\ \nonumber
&&-\frac{1}{3}(\gamma^{\pm})^2 \cos [ (\Omega_2-\Omega_3) t]-\frac{1}{6} (\gamma^{\pm})^2 \cos [(\Omega_2-\Omega_4) t]\\ 
&& +\frac{1}{36}\cos [(\Omega_3-\Omega_4) t]+\frac{5}{144}, 
\end{eqnarray}

\noindent We see that for our system the elements $\rho_{11}^{\pm}$ and $\rho^{\pm}_{44}$ are equal. That is, $\rho_{11}^{\pm} = \rho^{\pm}_{44}$.

%\begin{widetext}
\begin{eqnarray}\nonumber
\rho^{\pm}_{22}  &=&  (\alpha^{\pm})^4 (\eta^{\pm})^2+(\alpha^{\pm})^4+(\gamma^{\pm})^4  (\lambda^{\pm})^2+(\gamma^{\pm})^4 \\ \nonumber 
&& +2 (\alpha^{\pm})^2 (\gamma^{\pm})^2 (\eta^{\pm}  \lambda^{\pm} +1) \cos [(\Omega_1-\Omega_2) t]\\ \nonumber
&&+ \frac{1}{3} (\alpha^{\pm})^2 \left\lbrace \cos [(\Omega_1-\Omega_3) t]  + \eta^{\pm}  \sin [ (\Omega_1-\Omega_3) t]   \right\rbrace \\ \nonumber 
&& +\frac{1}{6} (\alpha^{\pm})^2 \left\lbrace \cos [( \Omega_1-\Omega_4) t] - 2 \eta^{\pm} \sin [ (\Omega_1-\Omega_4) t]\right\rbrace \\ \nonumber
&&+(\alpha^{\pm})^2 \cos [( \Omega_1-\Omega_5) t]+\frac{1}{3} (\gamma^{\pm})^2 \left\lbrace \cos [( \Omega_2-\Omega_3) t]+ \lambda^{\pm} \sin [( \Omega_2-\Omega_3) t] \right\rbrace \\ \nonumber
&&+\frac{1}{6} (\gamma^{\pm})^2 \left\lbrace \cos [( \Omega_2-\Omega_4) t]  -2 \lambda^{\pm} \sin [( \Omega_2-\Omega_4) t] \right\rbrace \\ \nonumber
&& +(\gamma^{\pm})^2 \cos [( \Omega_2-\Omega_5) t]-\frac{1}{36} \cos [( \Omega_3 -\Omega_4) t ]+\frac{1}{6} \cos [( \Omega_3-\Omega_5) t]\\
&&+\frac{1}{12} \cos [( \Omega_4 -\Omega_5) t]+\frac{49}{144},
\end{eqnarray}

%\end{widetext}
%\begin{widetext}
\begin{eqnarray}\nonumber
 \rho^{\pm}_{33}  &= & (\alpha^{\pm})^4 (\eta^{\pm})^2+(\alpha^{\pm})^4+(\gamma^{\pm})^4 (\lambda^{\pm})^2+(\gamma^{\pm})^4\\ \nonumber 
&& +2 (\alpha^{\pm})^2 (\gamma^{\pm})^2 (\eta^{\pm} \lambda^{\pm}+1) \cos [(\Omega_1-\Omega_2) t]\\ \nonumber
&& +\frac{1}{3} (\alpha^{\pm})^2 \left\lbrace \cos [\Omega_1-\Omega_3) t]- \eta^{\pm}  \sin [ (\Omega_1-\Omega_3) t]\right\rbrace \\ \nonumber 
&& +\frac{1}{6} (\alpha^{\pm})^2 \left\lbrace \cos [( \Omega_1-\Omega_4) t]  + 2 \eta^{\pm} \sin [ (\Omega_1-\Omega_4) t]\right\rbrace\\ \nonumber
&& -(\alpha^{\pm})^2 \cos [( \Omega_1-\Omega_5) t]+\frac{1}{3} (\gamma^{\pm})^2 \left\lbrace \cos [( \Omega_2-\Omega_3) t]- \lambda^{\pm} \sin [( \Omega_2-\Omega_3) t]\right\rbrace \\ \nonumber
&& + \frac{1}{6} (\gamma^{\pm})^2 \left\lbrace \cos [( \Omega_2-\Omega_4) t]+ 2 \lambda^{\pm} \sin [( \Omega_2-\Omega_4) t] \right\rbrace \\ \nonumber
&& -(\gamma^{\pm})^2 \cos [( \Omega_2-\Omega_5) t]-\frac{1}{36} \cos [(\Omega_3-\Omega_4)t ]-\frac{1}{6} \cos [( \Omega_3-\Omega_5) t]\\ 
&& -\frac{1}{12} \cos [( \Omega_4 -\Omega_5) t]+\frac{49}{144},
\end{eqnarray}
%\end{widetext}
%\begin{widetext}
\begin{eqnarray} \nonumber
 \rho^{\pm}_{23}  &=&  -\frac{1}{36} \im \left\lbrace 72(\alpha^{\pm})^4 {\eta^{\pm}}+72 (\gamma^{\pm})^4 {\lambda^{\pm}}+72 (\alpha^{\pm})^2 (\gamma^{\pm})^2 ({\eta^{\pm}} +{\lambda^{\pm}}) \cos [(\Omega_1-\Omega_2) t] \right.\\ \nonumber
&& + 12 (\alpha^{\pm})^2 ({\eta^{\pm}}+\im) \cos [ (\Omega_1-\Omega_3) t]+6 (\alpha^{\pm})^2 (\eta^{\pm} - 2 \im) \cos [ (\Omega_1-\Omega_4) t]\\ \nonumber 
&& -36 \im (\alpha^{\pm})^2 \eta^{\pm} \sin [ (\Omega_1-\Omega_5) t] + 12 (\gamma^{\pm})^2 (\lambda^{\pm} + \im) \cos [ (\Omega_2-\Omega_3) t]\\ \nonumber
&&  +6 (\gamma^{\pm})^2 (\lambda^{\pm} - 2 \im ) \cos [ (\Omega_2-\Omega_4) t]-36 i (\gamma^{\pm})^2 {\lambda^{\pm}} \sin [ (\Omega_2-\Omega_5) t]  \\ 
&& \left. -\im \cos [ (\Omega_3-\Omega_4) t]+6
\sin [ (\Omega_3-\Omega_5) t]-6 \sin [ (\Omega_4-\Omega_5) t]+\im \right\rbrace,
\end{eqnarray}
%\end{widetext}
%\begin{widetext}
\begin{eqnarray}\nonumber
 \rho^{\pm}_{32}  &=& \frac{1}{36} \im \left\lbrace 72(\alpha^{\pm})^4 {\eta^{\pm}}+72 (\gamma^{\pm})^4 {\lambda^{\pm}}+72 (\alpha^{\pm})^2 (\gamma^{\pm})^2 ({\eta^{\pm}}+{\lambda^{\pm}}) \cos [ (\Omega_1-\Omega_2) t] \right.\\ \nonumber
&& + 12 (\alpha^{\pm})^2 ({\eta^{\pm}}-\im) \cos [ (\Omega_1-\Omega_3) t ]+6 (\alpha^{\pm})^2 (\eta^{\pm} +2 \im) \cos [ (\Omega_1-\Omega_4) t] \\ \nonumber
&& +36 \im (\alpha^{\pm})^2 \eta^{\pm} \sin [ (\Omega_1-\Omega_5) t] +12 (\gamma^{\pm})^2(\lambda^{\pm} - \im) \cos [ (\Omega_2-\Omega_3) t]\\ \nonumber
&&  +6 (\gamma^{\pm})^2 (\lambda^{\pm} +2 \im) \cos [ (\Omega_2-\Omega_4) t ]+36 \im (\gamma^{\pm})^2 {\lambda^{\pm}} \sin [ (\Omega_2-\Omega_5) t] \\ 
&& \left. +\im \cos [ (\Omega_3-\Omega_4) t]+6
\sin [ (\Omega_3-\Omega_5) t]-6 \sin [ (\Omega_4-\Omega_5) t]-\im\right\rbrace.
\end{eqnarray}
%\end{widetext}

\noindent The normalization of $\hat{\varrho}_{AB}(t)$ is indeed 1, i.e. 
\begin{align*}
\langle \hat{\varrho}_{AB}(t)\rangle&= \mathrm{Tr}(\hat{\varrho}_{AB}(t)) = \rho^{\pm}_{11} + \rho^{\pm}_{22} + \rho^{\pm}_{33} + \rho^{\pm}_{44} \\ 
& = 2 \rho^{\pm}_{11} + \rho^{\pm}_{22} + \rho^{\pm}_{33} \\ 
& =  1
\end{align*}

\section{Density matrices \texorpdfstring{$\hat\varrho_{X,AB}(t)$ and $\hat\varrho_{Z,AB}(t)$}{varrho\_(X,AB)(t) and varrho\_(Z,AB)(t)}}
%\section{Density matrices $\hat\varrho_{X,AB}(t)$ and $\hat\varrho_{Z,AB}(t)$}

For the computational basis $S=\lbrace |00\rangle , |10\rangle , |01\rangle , |11\rangle \rbrace$, given the expression (\ref{eq:rhoX}) and $\ket{\phi_{1,2}}=\frac{1}{\sqrt{2}}\left(\ket{0}_A\pm\ket{1}_A\right)$ and $\hat{I}_B = |0\rangle\langle0|_B + |1\rangle \langle 1|_B$, we obtain for $\hat\varrho_{X,AB}(t)$, keeping in mind that $\rho^{\pm}_{11}=\rho^{\pm}_{44}$:

\begin{align*}\label{RXABmatrix} \nonumber
 \hat\varrho_{X,AB}(t) & = \frac{1}{2}\left\lbrace(\rho^{\pm}_{11} + \rho^{\pm}_{22}) \big(|00 \rangle \langle00| + |10 \rangle \langle 10 |\big) \right.\\ 
& + (\rho^{\pm}_{33} +\rho^{\pm}_{11}) \big(|01 \rangle \langle 01| + |11 \rangle \langle 11 |\big)   \\ \nonumber
& + \rho^{\pm}_{23} \big( |00 \rangle\langle	11| + | 10 \rangle \langle 01 |  \big)  \\ \nonumber 
& \left. + \rho^{\pm}_{32} \big( | 11 \rangle \langle 00 | + |01\rangle \langle 10 |\big) \right\rbrace ,\\ \nonumber
\end{align*}
or in matrix form 
\begin{eqnarray}\label{matrixform_rhoXAB}
\hat\varrho_{X,AB}(t)= 
\begin{bmatrix}
\frac{\rho^{\pm}_{11} + \rho^{\pm}_{22}}{2} & 0 & 0 &  \frac{\rho^{\pm}_{23}}{2}\\
0 & \frac{\rho^{\pm}_{11} + \rho^{\pm}_{22}}{2} &  \frac{\rho^{\pm}_{23}}{2} & 0\\
0 &  \frac{\rho^{\pm}_{32}}{2} &\frac{\rho^{\pm}_{33} + \rho^{\pm}_{11}}{2} & 0\\
\frac{\rho^{\pm}_{32}}{2} & 0 &0 & \frac{\rho^{\pm}_{33} + \rho^{\pm}_{11}}{2} 
\end{bmatrix}.\nonumber
\end{eqnarray}
\noindent For (\ref{eq:rhoZ})  and  $\ket{\psi_{1,2}}\equiv \ket{0}_A,\,\ket{1}_A$ we obtain the diagonal matrix
\begin{align*} \nonumber
 \hat\varrho_{Z,AB}(t)  &= \rho^{\pm}_{11}  |00 \rangle \langle00| + \rho^{\pm}_{22} |10 \rangle \langle 10 | \\ \nonumber 
& \quad +\rho^{\pm}_{33} |01 \rangle \langle 01| + \rho^{\pm}_{11} + |11 \rangle \langle 11 | ,\\ \nonumber
\end{align*}
\noindent or in matrix form 
\begin{eqnarray}\label{matrixform_rhoZAB}
\hat\varrho_{Z,AB}(t)= 
\begin{bmatrix}
\rho^{\pm}_{11}  & 0 & 0 &  0\\
0 & \rho^{\pm}_{22} & 0 & 0\\
0 & 0 & \rho^{\pm}_{33} & 0\\
0 & 0 &0 &  \rho^{\pm}_{11} \\ 
\end{bmatrix}.\nonumber
\end{eqnarray}
\noindent The reduced density matrices when I trace out the subsystem A from $\hat\varrho_{X,AB}(t)$ and $\hat\varrho_{Z,AB}(t)$ are given by
\begin{eqnarray}\label{ZB_reduced}
\hat\varrho_{B}(t)=\hat\varrho_{Z,B}(t)= \hat\varrho_{X,B}(t)=
\begin{bmatrix}
\rho^{\pm}_{11} + \rho^{\pm}_{22}  & 0 \\
 0 & \rho^{\pm}_{33} + \rho^{\pm}_{11} \\ 
\end{bmatrix}.\nonumber
\end{eqnarray}

\section{Explicit calculation for Quantum entropies} 

To write the entropies we can use the spectral decomposition for the above matrices. Given $\log\hat A=\sum\limits_n \ket{n}\bra{n}\log(A_n)$, the matrices $\hat\varrho_{B}(t)$, $\hat\varrho_{Z,B}(t)$ and $\hat\varrho_{X,B}(t)$ are already diagonal, hence the eigenvalues are the diagonal entries. We write 
\begin{eqnarray}\nonumber
y_1  = \rho^{\pm}_{11}(t) + \rho^{\pm}_{22}(t) &, \quad  y_2  = \rho^{\pm}_{33}(t) + \rho^{\pm}_{11}(t)\\ \nonumber
\end{eqnarray} 
\noindent The matrix $\hat\varrho_{Z,AB}(t)$ is also diagonal, so the eigenvalues are just the diagonal entries.
\begin{eqnarray}\nonumber
x^{Z}_1  = \rho^{\pm}_{11} &, \quad  x^{Z}_2  = \rho^{\pm}_{22} ,\quad 
x^{Z}_3  = \rho^{\pm}_{33} &, \quad  x^{Z}_4  = \rho^{\pm}_{11} \, .\\ \nonumber
\end{eqnarray}
\noindent For the eigenvalues of $\hat\varrho_{X,AB}(t)$, we solve $$\mathrm{det}(\hat\varrho_{X,AB}(t) - \lambda \mathbb{I}) =0 ,$$ 

\noindent We obtain the repeated eigenvalues 

\begin{align*}\nonumber
w_{1,2} & = \frac{1}{4} \left( 1 - \sqrt{1 - 4 \left( (\rho^{\pm}_{11} + \rho^{\pm}_{22}) (\rho^{\pm}_{33} + \rho^{\pm}_{11}) - |\rho_{23}^{\pm}|^2 \right)} \right) ,\\ \nonumber
w_{3,4} & = \frac{1}{4} \left( 1 + \sqrt{1 - 4 \left( (\rho^{\pm}_{11} + \rho^{\pm}_{22}) (\rho^{\pm}_{33} + \rho^{\pm}_{11}) - |\rho_{23}^{\pm}|^2 \right)} \right) .\\ \nonumber
\end{align*}
\noindent Given that $\rho^{\pm}_{11} = \rho^{\pm}_{44}$, the above eigenvalues simplify to the repeated eigenvalues :
\begin{align*}\nonumber
w^{X}_{1,2} & = \frac{1}{4} \left( 1 - \sqrt{(\rho^{\pm}_{22} - \rho^{\pm}_{33})^2 + 4 |\rho_{23}^{\pm}|^2} \right)\\ \nonumber
w^{X}_{3,4} & = \frac{1}{4} \left( 1 + \sqrt{(\rho^{\pm}_{22} - \rho^{\pm}_{33})^2 + 4 |\rho_{23}^{\pm}|^2} \right)\\ \nonumber
\end{align*}
\noindent For the eigenvalues of $\hat\varrho_{AB}(t)$, similarly we obtain 

\begin{align*}\nonumber
x_1  &= \rho^{\pm}_{11} \, , \, \,  x_4  = \rho^{\pm}_{44} =\rho^{\pm}_{11} ,\\ \nonumber
 x_{2} & = \frac{1}{2} \left( (\rho^{\pm}_{22} + \rho^{\pm}_{33}) - \sqrt{(\rho^{\pm}_{22} - \rho^{\pm}_{33})^2 + 4 |\rho_{23}^{\pm}|^2} \right) ,\\ \nonumber
 x_{3} &= \frac{1}{2} \left( (\rho^{\pm}_{22} + \rho^{\pm}_{33}) + \sqrt{(\rho^{\pm}_{22} - \rho^{\pm}_{33})^2 + 4 |\rho_{23}^{\pm}|^2} \right) .\\ \nonumber
\end{align*}

\noindent The entropies are given by:

\begin{eqnarray}\label{four spin case analytic expressions of entropies_appendix}
 S^\pm(X\vert B) & = & S^{\pm}(\hat\varrho_{X,AB}^\pm)-S^{\pm}(\hat\varrho^\pm_{X,B}),\\
 S^\pm(Z\vert B) & = & S^{\pm}(\hat\varrho_{Z,AB}^\pm)-S^{\pm}(\hat\varrho_{Z,B}^\pm),\\
 S^\pm(A\vert B) & = & S^{\pm}(\hat\varrho_{AB}^\pm)-S^{\pm}(\hat\varrho_B^\pm),
\end{eqnarray}

\noindent Then we obtain :

\begin{eqnarray}\label{four spin case in the explicit form_appendix}
 S^\pm(X\vert B)&=&-\sum\limits_{w^{X}_n\in[w^{X}_1,\ldots ,w^{X}_4]}w^{X}_n\log(w^{X}_n)
 + \sum\limits_{y_n\in[y_1,y_2]}y_n\log(y_n),\\
 S^\pm(Z\vert B)&=&-\sum\limits_{x^{Z}_n\in[x^{Z}_1,\ldots ,x^{Z}_4]}x^{Z}_n\log(x^{Z}_n)
 + \sum\limits_{y_n\in[y_1,y_2]}y_n\log(y_n),\\
 S^\pm(A\vert B)&=&-\sum\limits_{x_n\in[x_1,x_2,x_3,x_4]}x_n\log(x_n)
 + \sum\limits_{y_n\in[y_1,y_2,y_3,y_4]}y_n\log(y_n).
\end{eqnarray}

%\noindent The inequality $S(X|B)+ S(Z|B) \geq \mathrm{Log}_2(1/c) + S(A|B) ,$ equation(\ref{eq:BobsIgnorance}), becomes explicitly 
\noindent The LHS and RHS in (\ref{eq:BobsIgnorance}), become explicitly
\begin{eqnarray}\nonumber
\mathrm{LHS} &=&-\frac{1}{2} (1 - \sqrt{d})\mathrm{Log}_2(1 - \sqrt{d}) -\frac{1}{2} (1 + \sqrt{d})\mathrm{Log}_2(1 + \sqrt{d}) \\ 
&&- \rho^{\pm}_{22}\mathrm{Log}_2\rho^{\pm}_{22} -\rho^{\pm}_{33}\mathrm{Log}_2\rho^{\pm}_{33} - S(\rho_{B}(t)) , 
\end{eqnarray}
\begin{eqnarray}\nonumber
\mathrm{RHS}  &=& -1 + (\rho^{\pm}_{22} + \rho^{\pm}_{33})- \frac{1}{2} (\rho^{\pm}_{22} + \rho^{\pm}_{33} - \sqrt{d})\mathrm{Log}_2(\rho^{\pm}_{22} + \rho^{\pm}_{33} - \sqrt{d}) \\ 
&& - \frac{1}{2} (\rho^{\pm}_{22} + \rho^{\pm}_{33} + \sqrt{d})\mathrm{Log}_2(\rho^{\pm}_{22} + \rho^{\pm}_{33} + \sqrt{d}) ,
\end{eqnarray}
\\
\noindent where $d \equiv (\rho^{\pm}_{22} - \rho^{\pm}_{33})^2 + 4 |\rho_{23}^{\pm}|^2 $  and $\rho^{\pm}_{44} = \rho^{\pm}_{11}$ and 

\begin{eqnarray}\label{ZB_reduced_explicit}
&&\hat\varrho_{B}(t)=\hat\varrho_{Z,B}(t)= \hat\varrho_{X,B}(t)=
\begin{bmatrix}
\rho^{\pm}_{11} + \rho^{\pm}_{22}  & 0 \\
 0 & \rho^{\pm}_{33} + \rho^{\pm}_{11} \\ 
\end{bmatrix}.\nonumber
\end{eqnarray}

\noindent Hence, we have for the entropy of the marginal matrices

\begin{align*}
S^{\pm}(\hat\varrho^{\pm}_{B})&=  - (\rho^{\pm}_{11} + \rho^{\pm}_{22})\mathrm{Log}_2(\rho^{\pm}_{11} + \rho^{\pm}_{22}) \\
& \quad    - (\rho^{\pm}_{33} + \rho^{\pm}_{11})\mathrm{Log}_2(\rho^{\pm}_{33} + \rho^{\pm}_{11}) .
\end{align*}

%\section{First Section}

%\subsection{First Subsection}

%\subsubsection{First Sub Subsection}

%\threesubsection{First lowest-level subsection}

%\section{Conclusion}

% Experimental section

%\section{Experimental Section}
%\threesubsection{First part of experimental section}\\
%\threesubsection{Second part of experimental section}\\

% Acknowledgements
%\medskip
%\textbf{Acknowledgements} \par %delete if not applicable))
%Please insert your acknowledgements here

% References
\medskip

% Use the following code if you wish to generate your bibliography with BibTeX;
% replace the string "MSP-template" below with the name(s) of
% the BibTeX data base(s) you want to use.
% The resulting bibliography-output (the content of the .bbl file)
% must be pasted back into this file before submission.
% Please also include your BibTeX data base file(s) in your submission
% so that we can re-run BibTeX if necessary.
%
%\bibliographystyle{MSP}
%\bibliography{bibliography_new}

% Table of contents entry should be 50 - 60 words long
% Image should be 55 mm broad and 50 mm high or 110 mm broad and 20 mm high

%\begin{figure}
%\textbf{Table of Contents}\\
%\medskip
%  \includegraphics{toc-image.png}
%  \medskip
%  \caption*{ToC Entry}
%\end{figure}

\end{document}